\documentclass[a4paper,11pt]{article}
\usepackage[T1]{fontenc}
\usepackage{lmodern}
\usepackage[usenames,dvipsnames,svgnames,table,HTML]{xcolor}
\usepackage{jheppub}
\usepackage{orcidlink}
\usepackage{aligned-overset}
\usepackage{tikz}
\usepackage{tikz-feynman}
\usetikzlibrary{calc}
\usepackage{multirow}
\usepackage{enumerate}
\usepackage{amsmath}
\usepackage{pgf}
\usepackage{subcaption}
\usepackage{import}

\graphicspath{{figures/}}

\renewcommand{\d}{\mathrm{d}}

\renewcommand{\O}{\mathcal{O}}
\newcommand{\Mpl}{M_\mathrm{Pl}}
\newcommand{\RN}[1]{%
  \textup{\uppercase\expandafter{\romannumeral#1}}%
}

\preprint{\vbox{\hbox{LMU-ASC 03/26}}}

\title{\boldmath Inflationary Bispectra and IR Physics from Quantum Simulators}

\author{Matthias Nowinski\orcidlink{0009-0006-4736-6976}}
\author{and Ivo Sachs\orcidlink{0000-0003-2526-8655}} 

\affiliation{Arnold-Sommerfeld-Center for Theoretical Physics, Ludwig-Maximilians-Universit\"at M\"unchen, Theresienstr. 37, D-80333 Munich, Germany}

\emailAdd{m.nowinski@lmu.de}

\abstract{Bose-Einstein condensates have been used  successfully as quantum simulators for early-universe particle production. Here we extend this analogy to the interacting theory. To this end, we compute the leading interaction terms expected in a general experimental setup that determine the 3-point correlations. We apply them to cosmologically motivated expanding model universes, but also periodically oscillating ones. While expanding scale factors do not necessarily lead to easily observable signatures due to experimental constraints on the realizable scale factor, periodically driven ones produce resonant correlators that grow linearly over time and lend themselves well to experimental detection. One direct application of this analysis is the experimental study of late-time infrared divergencies often discussed in the context of massless scalars in de Sitter space.}

\begin{document}
\maketitle
\flushbottom

\section{Introduction}
In spite of considerable progress in the theoretical treatment of QFT in curved spacetime, due in parts to its cosmological applications \cite{Mukhanov:1981xt}, the observations have not quite caught up to predictions yet \cite{Planck:2018nkj}. A very promising approach to bridge the gap between theory and observations is the use of quantum simulators. See for example \cite{Barcelo:2005fc, Schutzhold:2025qna} for reviews on this topic. While the original idea of simulating QFT in curved spacetime phenomena using some more accessible model system is almost as old as the original theory itself \cite{Unruh:1980cg}, very promising advances have been made in the recent past, thanks to more advanced technology and a steady push of the experimental Bose-Einstein condensate (BEC) community. See, for example, \cite{PhysRevA.106.033313, Viermann:2022wgw, PhysRevX.8.021021,Jain:2007gg, Sparn:2024vta, Schmidt:2024zpg, ViermannPHD} for an overview of recent developments.

Despite the recent progress, however, mostly phenomena of the free theory like particle production and mode red-shift have been investigated experimentally so far. To the best of our knowledge, there are no works on predicting or measuring self-interactions of BEC simulators analog for QFT in curved spacetime. Coming from the theory side, we are very interested in the behavior of the theory beyond leading order. Therefore, we apply the methods outlined in \cite{Viermann:2022wgw} and derive the self-interactions of the perturbations around the background BEC in section \ref{sec:theory}. 

Using a phase-modulus split for the BEC action (see \eqref{eq:scale_free_action_final}), we find that the interaction terms have the shape $\mathcal{L}_\mathrm{int} = \sqrt{-g} V(\phi, \dot\phi, \vec\nabla\phi)$, where $\phi$ is the quantum field and $g$ the determinant of the analog space-time metric. This is very similar to the action of prominent phenomenological models for inflation, and we show that the interaction terms of the BEC do closely resemble some of the interactions derived in the effective field theory (EFT) of inflation \cite{Cheung:2007st}. This potentially allows simulation of a phenomenologically relevant model of the early universe.

Measuring interactions in an analog QFT in curved spacetime at all would already constitute a significant test of the methods used in the field, more precisely the application of the Schwinger-Keldysh formalism, as well as an important step in the experimental study of these systems. We assume a simulated Friedmann–Lemaître–Robertson-Walker (FLRW) universe, and present first predictions of the momentum space 3-point function (also called bispectrum) of the density and phase fluctuations of the BEC in section \ref{sec:growing_a}.

From the theory point of view, the most interesting aspect of this study is the late time, or IR, physics. It is known that a massless, minimally coupled scalar field in de Sitter space exhibits perturbative IR divergences. This is noticeable through so-called secular growth of the late-time correlation functions in the perturbative expansion. There have been many works on this problem in the recent past, see for example \cite{Beneke:2012kn, Gorbenko:2019rza, Beneke:2023wmt}. However, there still seems to be some debate about this topic, as some authors are not quite satisfied with the broadly accepted narrative that some kind of EFT is necessary to tame the IR divergences. One example for this claim is \cite{Palma:2025oux}\footnote{A central claim of that paper is that fields with only derivative interactions do not exhibit IR divergences in de Sitter. Our analysis shows such divergences, albeit for polynomial scale factors, in table \ref{tab:late_time_3pt_functions}.}, which proposes an alternate way of dealing with IR problems. We show in section \ref{sec:growing_a} that these IR problems are generally also present in the quantum simulator, for many different choices of scale factor $a(t)$. Although they are noticeable only at late times, the large degree of control in the experiment might allow a precise investigation into the nature of these phenomena.

Additionally, in section \ref{sec:periodic_a}, we present a slightly different setup with a periodic scale factor. This is motivated less by actual cosmology (even though there are some cyclic universe models being proposed \cite{Steinhardt:2004gk}), and more by the attempt to find a parameter space where the interaction terms are as strong as possible, so that experimental detection is easiest. We find that, as naively expected, there are resonant growth conditions for both the 2- and 3-point functions. Crucially, it seems possible to separate the two by choice of experimental parameters, such that resonant interaction contributions can be measured, in a setting where resonant particle production is suppressed.

\section{BEC Quantum Simulators} \label{sec:theory}

We follow the general theoretical framework laid out in \cite{PhysRevA.106.033313} for the quantum simulator. The setup is a BEC of some cold atom gas caught in an optical trap and confined to $d=2$ space dimensions. This gas has an effective quartic self-interaction with strength $\lambda = \lambda(t,\vec x)$ that can be tuned using Feshbach resonances and therefore can depend on space and time. Since we are mostly interested in time-dependent phenomena and not spatially curved background metrics, we will soon specialize to a spatially constant background, which is a good approximation at the center of the experimental setup.

\subsection{Action}

Our starting point differs slightly from \cite{PhysRevA.106.033313} since we model this gas as a relativistic (as opposed to non-relativistic) scalar $\Phi$ with charge $q$, coupled to a classical electromagnetic field $A_\mu$. This description is more tailored towards the high-energy physics community. Since we want to be able to extract the non-relativistic limit manifestly later on, we keep $c$ and $\hbar$ explicit for now. Using Cartesian coordinates $\vec x = (x,y)$ and laboratory time $t$ with the Minkowski metric $(-, +, +)$, the action is
\begin{align}
    S = \int \d t \d x \d y \Bigg( - \big( D_\mu \Phi\big)^\dagger\big( D^\mu \Phi \big) - \frac{m^2c^2}{\hbar^2} \Phi^\dagger \Phi - \frac{\lambda}{4} \big( \Phi^\dagger \Phi \big)^2 \Bigg). \label{eq:starting_action}
\end{align}
Here, $D_\mu = \partial_\mu + iq/(\hbar c) A_\mu$ is the usual covariant derivative. We expand this field around some classical, complex-valued BEC background. For this, we split the full field $\Phi$ radially using a real background phase $S_0$ with a fluctuation field $\sigma$ and a real modulus $R_0$ with another fluctuation field $r$ as
\begin{align}
    \Phi = e^{i (S_0 + \sigma)} \Bigg( R_0 + r \Bigg). \label{eq:field_expansion}
\end{align}
This ansatz is different from that used in the related literature \cite{PhysRevA.106.033313}, but it will turn out to be much better suited for treating interaction terms later on. From here, we can compute the classical equations of motion (EOM) for the background fields $R_0$ and $S_0$, which turn out to be
\begin{subequations} \label{eq:euler_eq}
\begin{align}
    0 &= 2 \partial_\mu \Big( \bar W^\mu R_0^2 \Big), \\
    0 &= -2 R_0 \bar W_\mu \bar W^\mu + 2 \underbrace{\Box R_0}_{\approx \, 0} - 2 \frac{m^2c^2}{\hbar^2} R_0 - \lambda R_0^3. \label{eq:rho0_eom}
\end{align}
\end{subequations}
We have defined the vector $W_\mu$ with the background part $\bar W_\mu$ as follows
\begin{subequations}
\begin{align}
    W_\mu &= \partial_\mu \Big( S_0 + \sigma \Big) + \frac{q}{\hbar c} A_\mu, \\
    \bar W_\mu &= \partial_\mu S_0  + \frac{q}{\hbar c} A_\mu,
\end{align}
\end{subequations}
This is the point were we specialize to a constant background density $R_0^2$, so we drop the $\Box R_0$ term in \eqref{eq:euler_eq}, which is essentially the "quantum pressure" term that is also dropped in similar studies \cite{PhysRevA.106.033313}. The equations \eqref{eq:euler_eq} are often called Madelung or Euler equations.
From here, we can also easily read off the non-relativistic limit, where we assume that the overall phase $e^{i S_0}$ is dominated by the mass (rest energy) plus the chemical potential $\mu_0(t) = \partial_t M(t)$ and write
\begin{align}
    S_0 = - \frac{mc^2}{\hbar} t - \frac{M(t)}{\hbar} + \O(c^0) \quad \mathrm{and} \quad A_x = 0, \quad A_y &= 0, \quad A_t = \frac{V(t)}{c}.
\end{align}
Afterwards, we drop all terms suppressed by $1/c$ -- in particular the $\partial_t / c$ term hidden inside every contraction of $\partial_\mu$. We can then express the action \eqref{eq:starting_action} in terms of the fluctuation fields $\sigma$ and $r$, and simplify using the background EOM \eqref{eq:rho0_eom} to find
\begin{align}
    S = \int \d t \d x \d y \Bigg[ &- \frac{2m R_0^2}{\hbar} \dot\sigma - \frac{2m}{\hbar} \dot\sigma \Big( 2 R_0 r + r^2 \Big) + R_0^2 \Big( \vec\nabla \sigma \Big)^2 - \Big( \vec\nabla r \Big)^2 \notag\\
    &- \Big( \vec \nabla \sigma \Big)^2 \Big( 2 R_0 r + r^2 \Big) - \frac{\lambda}{4} \Big(2 R_0 r + r^2\Big)^2 \Bigg]. \label{eq:parametrized_lagrangian}
\end{align}
The first term is linear in the fluctuations and a total derivative, so we will drop it from now on\footnote{Actually, boundary terms do play an important role in the computation of correlation functions \cite{Arroja:2011yj}. However, this particular term is a tadpole and will only contribute for $\vec k = 0$. Away from this background contribution, we can ignore it exactly.}. The final simplification is dropping the $(\vec \nabla r)^2$ term, assuming that the effective mass of the density fluctuations is much bigger than the kinetic energies we expect. This is justified when we keep to lower momenta  in accordance with \cite{PhysRevA.106.033313}. 

Now, we can see that $r$ only appears in \eqref{eq:parametrized_lagrangian} in the combination $\delta\rho \coloneq 2 R_0 r + r^2$, so we can perform the non-linear field redefinition to switch to this variable. This also has a straightforward physical interpretation, since 
\begin{align}
    \rho = \Phi^\dagger \Phi = \rho_0 + \delta\rho,
\end{align}
so we are now working directly with the density and phase fluctuations $\delta \rho$ and $\sigma$ of the BEC field.

As a result, the action is now quadratic in $\delta\rho$ and does not contain any derivatives of it, so we can exactly integrate it out via its EOM
\begin{align}
    \delta \rho = - \frac{2}{\lambda} \Bigg( \frac{2m}{\hbar} \dot\sigma + \Big( \vec\nabla \sigma \Big)^2 \Bigg). \label{eq:deltarho_eom_solution}
\end{align}
Inserting this back into the action \eqref{eq:parametrized_lagrangian}, we find the very simple and (up to higher momentum contributions) exact asction
\begin{align}
    S = \int \d t \d x \d y \Bigg[ \frac{1}{\lambda} \Bigg( \frac{2m}{\hbar} \dot\sigma + \Big( \vec\nabla \sigma \Big)^2 \Bigg)^2 - \ R_0^2 \Big( \vec\nabla \sigma \Big)^2 \Bigg].
\end{align}
To go to a canonical normalization and see that this is indeed the action of a massless scalar field in a FLRW universe, we define
\begin{align}
    \phi \coloneq \sqrt{2 \rho_0} \sigma, \quad a^2(t) \coloneq \frac{4 m^2}{\hbar^2 \rho_0} \frac{1}{\lambda(t)}, \quad \kappa \coloneq \frac{\hbar}{m \sqrt{2\rho_0}},
\end{align}
and expand the quadratic to find
\begin{align}
    S = \int \d t \d x \d y \Bigg[ \frac{a^2(t)}{2} \dot\phi^2 - \frac{1}{2} \Big( \vec\nabla \phi \Big)^2 + \frac{\kappa}{2} a^2(t) \dot\phi \Big( \vec\nabla \phi \Big)^2 + \frac{\kappa^2}{8} a^2(t) \Big( \vec\nabla \phi \Big)^4 \Bigg]. \label{eq:expanded_full_action}
\end{align}
This can be identified as
\begin{align}
    S &= \frac{1}{2} \int \d t \d x \d y \, a^2(t) \Bigg( -g^{00} \partial_t \phi_2 \partial_t \phi_2 - g^{jk} \partial_j \phi_2 \partial_k \phi_2 \Bigg) + S_\mathrm{int}
\end{align}
using the metric
\begin{align}
    g_{\mu\nu} &= \begin{pmatrix}
        -1 & 0 \\
        0 & a^2(t) \delta_{jk}
    \end{pmatrix}
\end{align}
Up to a re-scaling of the fields $\Phi$, the quadratic action agrees with \cite{PhysRevA.106.033313}. The new part are the two interaction terms
\begin{align}
    L_\mathrm{int} =  \frac{\kappa}{2} a^2(t) \dot\phi \Big( \vec\nabla \phi \Big)^2 + \frac{\kappa^2}{8} a^2(t) \Big( \vec\nabla \phi \Big)^4.
\end{align}
They have the correct $a(t)$-dependence, since they naturally come with $\sqrt{g} = a^2(t)$, and feature only shift-symmetric derivatives, which is expected since the field $\phi$ is the phase of the BEC. 

Up to the power of $a(t)$ in front, both interaction terms are also featured in the natural model for the actual cosmological inflation, the so-called EFT of inflation \cite{Cheung:2007st}, as can be read off from the Lagrangian \eqref{eq:dSEFT_action_final}. This allows us to study a phenomenologically relevant model for inflation in the lab.

\medskip

So far, it is not very easy to read off if the interactions are large or small, since the coupling $g$ is dimensionful. Therefore, we transition to dimensionless quantities 
\begin{align}
    t \xlongrightarrow{\mathrm{replace}} \tau t, \quad x \xlongrightarrow{\mathrm{replace}} \xi x, \quad y \xlongrightarrow{\mathrm{replace}} \xi y, \quad \phi \xlongrightarrow{\mathrm{replace}} Z \phi.
\end{align}
For this, we use the characteristic scales of the problem, which are given by the speed of sound $c_s$ and the healing length $\xi$. Together, they give the characteristic time scale $\tau$. Note that $c_s$, $\xi$ and $\tau$ are all time-dependent, because $\lambda$ and therefore the scale factor $a(t)$ are. For estimating of the size of the interactions, we will use values at the onset of expansion, which we denote by adding a index "$\mathrm{i}$". The speed of sound can be read off from \eqref{eq:expanded_full_action} as the coefficient in front of the time derivative term, and $\xi$ is a natural combination of the constants of the problem
\begin{align}
    c_s^{\mathrm{i}} = \frac{1}{a_\mathrm{i}} = \frac{\hbar \sqrt{\lambda_{\mathrm{i}} \rho_0}}{2m}, \quad \xi_{\mathrm{i}} = \frac{\hbar}{m c_s^{\mathrm{i}}}, \quad \tau_{\mathrm{i}} = \frac{\xi_{\mathrm{i}}}{c_s^{\mathrm{i}}}.
\end{align}
Then, demanding that the action $S$ in \eqref{eq:expanded_full_action} is proportional to $\hbar$, we find that
\begin{align}
    Z_{\mathrm{i}} = \sqrt{\frac{c_s^{\mathrm{i}} \hbar}{\xi_{\mathrm{i}}}}.
\end{align}
Finally, using the normalized scale factor $\tilde{a}(t) = a(t)/a_\mathrm{i}$ with initial condition $\tilde{a}(t) = 1$ for early times and renaming 
\begin{align}
    \tilde{a}(t) \xrightarrow{\mathrm{rename}} a(t), \quad \frac{\kappa Z_\mathrm{i}}{\tau_\mathrm{i} (c_s^\mathrm{i})^2}\xrightarrow{\mathrm{rename}} \kappa,
\end{align}
we can write down the full scale-free action as
\begin{align}
    S = \hbar \int \d t \d x \d y \Bigg[ \frac{a^2(t)}{2} \Bigg( \dot\phi + \frac{\kappa}{2} \Big( \vec\nabla \phi \Big)^2 \Bigg)^2 - \  \Big( \vec\nabla \phi \Big)^2 \Bigg]. \label{eq:scale_free_action_final}
\end{align}
The single parameter that controls the size of the interactions now is $\kappa$, which is a dimensionless number. To get an estimate for the experimental size of this number, we use the mass of potassium-39 $m \approx 6.5 \cdot 10^{-26} \, \mathrm{kg}$, the reduced Planck's constant $\hbar \approx 1.1 \cdot 10^{-34} \, \mathrm{Js}$ and $\lambda_\mathrm{i} \approx 7.7 \cdot 10^{41} \, \mathrm{J^{-1} m^{-2}}$ as the initial coupling\footnote{The large number in front of the unit might look concerning. However, the dimensionless coupling is $\lambda_\mathrm{i} \hbar \xi\mathrm{i}^2 / \tau = \kappa \approx 0.13$, which clearly is in the perturbative range.}. The estimate for $\lambda$ is computed from eq (2.7) in \cite{ViermannPHD}, using an initial $a_s^\mathrm{i} = 50 a_\mathrm{B} \approx 2.7 \cdot 10^{-9}\, \mathrm{m}$ ($a_s^{(\mathrm{i})}$ is called the (initial) scattering length and $a_\mathrm{B} \approx 5.3 \cdot 10^{-11} \, \mathrm{m}$ is the Bohr radius) and a $\omega_z = 2\pi \cdot 1.6 \, \mathrm{kHz}$. Both experimental parameters are also taken from \cite{ViermannPHD}. Note that the $\lambda$ found in references \cite{ViermannPHD, PhysRevA.106.033313} is using a rescaled non-relativistic field $\xi$ with $\phi = \hbar/\sqrt{2m} \cdot \xi$, which necessitates a rescaling of the experimental values for $\lambda$ and $\rho_0$. Finally, we find
\begin{align}
    \kappa \approx 0.133,
\end{align}
which is clearly in the perturbative range but also not extremely small, so that we might hope to see its effects in actual experiments.

\subsection{Observables} \label{sec:observables}

Now that we have a theory, the most important question is what the observables are. Since the experimental setup is a 2-dimensional BEC, the orthogonal third dimension can be used to image the density via optical means, but also to store other moving parts or fields. There are 3 main routes for observables:
\begin{enumerate}
    \item Direct imaging of density correlations
    \item Imaging of momentum space density correlations via "phase space rotation" \cite{PhysRevA.90.043611, Liebster:2023mmf}
    \item Homodyne measurement techniques inspired from quantum optics that image the phase of the condensate \cite{PhysRevLett.128.250402}
\end{enumerate}

\subsubsection*{Direct Density Measurements via $\delta\rho$} 

Let us first focus on the first avenue. The observables here are momentum space correlation functions of the normalized density contrast $(\rho - \rho_0) / \rho_0 = \delta\rho / \rho_0$. For this, a picture is taken of the condensate vertically to the 2-dimensional plane it is confined to and the density contrast is inferred from the picture. Then, a Fourier transform is performed and correlation functions are measured in Fourier space.

To relate those correlation functions to objects that our theory can predict, we first have to relate the density contrast to the dimensionless fluctuating fields via \eqref{eq:deltarho_eom_solution}
\begin{align}
    \frac{\delta\rho}{\rho_0} = - \kappa \, a^2(t) \Bigg( \dot\phi + \frac{\kappa}{2} \Big( \vec\nabla \phi \Big)^2 \Bigg). \label{eq:nonlinear_deltarho}
\end{align}
This is a composite operator in the quantum theory and one might expect that this leads to non-trivial contributions from higher field correlation functions. However, we have to remember that the above operator identity holds in the Heisenberg picture, which we shall denote by an index $\mathrm{H}$ like $\phi_\mathrm{H}(t)$, while we compute observables in the interaction picture, where we write the index $\mathrm{I}$. While the transition is trivial for the fields themselves
\begin{align}
    \phi_\mathrm{H} = U^\dagger \phi_\mathrm{I} U, \quad \mathrm{where} \quad U = T e^{- i \int \d t H_\mathrm{I}},
\end{align}
it is more complicated for the time derivatives. A quick computation shows
\begin{align}
    \dot\phi_\mathrm{H} = \dot U^\dagger \phi_\mathrm{I} U + U^\dagger \phi_\mathrm{I} \dot U + U^\dagger \dot\phi_\mathrm{I} U = U^\dagger \Big( \dot\phi_\mathrm{I} + i \big[ H_\mathrm{I}, \phi_\mathrm{I}\big] \Big) U. \label{eq:heisenberg_interaction_relation}
\end{align}
The Hamiltonian of our system can be found via Legendre transform to be
\begin{align}
    H = \frac{p^2}{2 a^2} - \Big( \vec\nabla \phi \Big)^2 - \frac{\kappa}{2} p \Big( \vec\nabla \phi \Big)^2,
\end{align}
where 
\begin{align}
    p = a^2 \Bigg(\dot\phi + \frac{\kappa}{2} \Big( \vec\nabla \phi \Big)^2 \Bigg).
\end{align}
Using the canonical commutation relations, \eqref{eq:heisenberg_interaction_relation} simplifies to
\begin{align}
    \dot\phi_\mathrm{H} = U^\dagger \Bigg( \dot\phi_\mathrm{I} - \frac{\kappa}{2} \Big( \vec\nabla \phi_\mathrm{I} \Big)^2 \Bigg) U. \label{eq:heisenberg_interaction_relation_final}
\end{align}
In total, the effect is that
\begin{align}
    \frac{\delta\rho_\mathrm{I}}{\rho_0} = - \kappa \dot\phi_\mathrm{I},
\end{align}
since the extra piece in \eqref{eq:nonlinear_deltarho} exactly cancels the additional term in \eqref{eq:heisenberg_interaction_relation_final}.

Any correlation function will be proportional to $(2\pi)^2 \delta^{(2)}(\vec k_1 + \cdots + \vec k_n)$, assuming that boundary effects in the BEC are negligible and the BEC is very large compared to the wavelengths. We denote the correlator after removing this universal factor using a prime as $\langle \O \rangle^\prime$. Correlators of the normalized density-contrast can then be written using
\begin{align}
    B_\rho^n\Big(\vec k_1, \vec k_2, \ldots, \vec k_n\Big) \coloneq \frac{1}{\rho_0^n} \Big\langle \delta\rho_{\vec k_1} \delta\rho_{\vec k_2} \cdots \delta\rho_{\vec k_n} \Big\rangle^\prime = \big( -\kappa \big)^n a^{2n}(t) \Big\langle \dot\phi^\mathrm{I}_{\vec k_1} \dot\phi^\mathrm{I}_{\vec k_2} \cdots \dot\phi^\mathrm{I}_{\vec k_n} \Big\rangle^\prime.
\end{align}
The right hand side can then be computed using the Schwinger-Keldysh formalism as will be explained in the next section.

The second avenue of measurements via "phase space rotation" can be used in BEC systems like the one we are using to directly image the momentum space correlation functions without having to go through position space in the data analysis \cite{PhysRevA.90.043611, Liebster:2023mmf}. However, this does not offer any conceptually new, so we will not go into further detail.

\subsubsection*{Homodyne Phase Measurements}

An orthogonal approach to observables are so-called homodyne or heterodyne techniques. They can be used to measure correlation functions of the phase fluctuations of a BEC \cite{PhysRevLett.128.250402}, and thus can be used to directly access correlation functions of the field $\phi$, instead of having to go through its time derivative like in the direct density correlators. This can be done by storing multiple 2d BEC sheets next to each other and then overlapping them. In terms of our dimensionless field $\phi$, we have
\begin{align}
    \sigma = \frac{1}{\sqrt{2\rho_0}} Z \phi = \kappa \phi.
\end{align}
Therefore, the homodyne observables are
\begin{align}
    B_S^n\Big(\vec k_1, \vec k_2, \ldots, \vec k_3\Big) \coloneq \Big\langle \sigma_{\vec k_1} \sigma_{\vec k_2} \cdots \sigma_{\vec k_3} \Big\rangle^\prime = \kappa^n \Big\langle \phi_{\vec k_1} \phi_{\vec k_2} \cdots \phi_{\vec k_3} \Big\rangle^\prime.
\end{align}
For the rest of this paper, we will only work with the field $\phi$. We will mostly not refer to any experimental parameters and work with the dimensionless quantities to keep the following discussion self-contained. It is understood that any time is measured in units of $\tau_\mathrm{i}$, any length in units of $\xi_\mathrm{i}$ and so on. Lastly, momenta $k$ are understood to be in units of $(\mathrm{time})^{-1}$, converted via the speed of sound $c_s$.

\section{Growing Scale Factors} \label{sec:growing_a}

In the following, we will consider two interaction terms in a (analog) FLRW universe background:
\begin{align}
    L_\mathrm{int} \supset \quad +\frac{\lambda}{3!} a^2(t) \phi^3, \qquad +\frac{\kappa}{2} \, a^2(t) \dot\phi \Big( \vec\nabla \phi \Big)^2. \label{eq:vertices}
\end{align}
The first vertex, while not present in our setup, is the  default coupling any high-energy physicist  would write down (up to a sign). It will be used for clarifications about the formalism and for comparison with the high-energy literature. The second vertex is the three-field interaction term that appears in the quantum simulator setup \eqref{eq:scale_free_action_final}. Since 4-point functions are more elusive than 3-point functions, we will not consider the 4 field interaction vertex here.

As a first application of these interactions, we want to consider cosmologically relevant models, i.e.\ FLRW universes with an expanding scale factor $a(t)$. The most useful scale factors are
\begin{enumerate}
    \item $a(t) = e^{H t}$ exponential expansion (de Sitter space)
    \item $a(t) = Q \, t^\gamma, \gamma > 1$ accelerating polynomial
    \item $a(t) = Q \, t^\gamma, \gamma < 1$ decelerating polynomial (matter and radiation domination)
\end{enumerate}
In order to predict any correlation function, we use the Schwinger-Keldysh formalism (see for example \cite{Weinberg:2005vy, baumann_lecture_notes} for an introduction). At leading order, we only need the Wightman function as a propagator, which is given by a product of mode functions $\phi_k(t)$ without time-ordering
\begin{align}
    G_k(t, t') = \phi_k^*(t) \phi_k(t'). \label{eq:propagator}
\end{align}
Momentum space correlators, for example the 3-point function using the $\lambda$ interaction in \eqref{eq:vertices}, can then be computed by integrating the internal time $t'$ over the experimental duration while fixing the external time $t$ at the measurement time $t_\mathrm{m}$, and then taking the imaginary part:
\begin{align}
    \langle \phi_{k_1}(t_\mathrm{m}) \phi_{k_2}(t_\mathrm{m}) \phi_{k_3}(t_\mathrm{m}) \rangle = - 2 \lambda \, \mathrm{Im} \int\limits_{t_0}^{t_\mathrm{m}} \d t \ a^2(t) G_{k_1}(t_\mathrm{m}, t) G_{k_2}(t_\mathrm{m}, t) G_{k_3}(t_\mathrm{m}, t). \label{eq:schwinger_keldysh_example}
\end{align}
If we want to use the other interaction vertex in \eqref{eq:vertices}, we let the contained time derivative act on the $t$ argument of the propagators under the time integral and replace the spatial derivative $\vec\nabla$ by the momentum $- i \vec k_j$ of the propagator it is acting on. Then, as usual, we have to sum all possible contractions and put the correct coupling constant in front.

The last step is projecting onto the correct initial state. If we assume the system to be in a vacuum state at early times we have to send $t_0 \longrightarrow - \infty (1 \mp i \epsilon) \eqcolon -\infty^\mp$ with some $\epsilon > 0$. The sign depends on which branch of the Schwinger-Keldysh contour we are on. For our purposes, we can choose the sign such that the resulting expression is finite. For details, refer to \cite{Chen:2017ryl, baumann_lecture_notes}. This is similar to the $i \epsilon$ prescription in flat space scattering computations and makes sure that early times do not contribute strongly to the correlation function.

A vacuum state is not necessarily realistic \cite{Sanchez-Kuntz:2022gds}, but sufficient to study the most important phenomena. Therefore, we will restrict our discussion in this paper mostly to this case. To include finite temperature effects, we would simply replace the propagator according to
\begin{align}
    G_k(t, t') = \phi_k^*(t') \phi_k(t) \longrightarrow (1 + n_k) \phi_k^*(t') \phi_k(t) + n_k \phi_k(t') \phi_k^*(t),
\end{align}
where $n_k = \big( e^{\hbar k / (k_\mathrm{B} T)} - 1 \big)^{-1}$ using Boltzmann's constant $k_\mathrm{B}$ and the initial temperature $T$.

There is one more amendment for correlators of the derivative field $\dot\phi$, which are important for the directly observable density correlation functions of the BEC, as per section \ref{sec:observables}. In order to compute these, we simply act with a time derivative on the $t_\mathrm{m}$ argument of the propagators in \eqref{eq:schwinger_keldysh_example}.

\subsection{Mode Functions}

\begin{figure}
    \centering
    \includegraphics{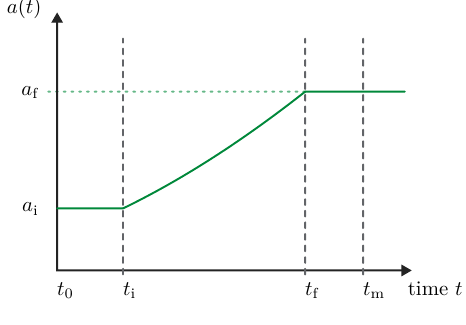}
    \caption{Scale factor $a(t)$ as a function of time. The experiment is set up at time $t_0$. Between the times $t_\mathrm{i}$ and $t_\mathrm{f}$, the scale factor is continuously interpolating the values $a_\mathrm{i}$ and $a_\mathrm{f}$. The correlation function is then measured at time $t_\mathrm{m}$, when the scale factor has been held at a constant value for some time. This figure is inspired by \cite{Sanchez-Kuntz:2022gds}.}
    \label{fig:scale_factor}
\end{figure}

The mode functions $\phi_k(t)$ in \eqref{eq:propagator} are solutions to the free EOM of the classical field, which in this case is
\begin{align}
    \ddot \phi_k(t) + d \frac{\dot a(t)}{a(t)} \dot\phi_k(t) + \frac{k^2}{a^2(t)} \phi_k(t) = 0. \label{eq:general_EOM}
\end{align}
This is a second order differential equation, so there are two solutions in general superposition
\begin{align}
    \phi_k(t) = C_k \psi_k(t) + D_k \mu_k(t).
\end{align}
Their exact shape depends on the scale factor $a(t)$ and the coefficients $C_k$ and $D_k$ are determined by the initial conditions and the Wronskian condition. 

Experimentally, we cannot extrapolate the polynomial nor the exponential scale factors to the infinite past or future, since there is only a finite range of accessible scale factors, which are constrained to be of order $a(t) \overset{!}{=} \O(1)$. For a realistic computation, we are using the rough experimental setup outlined in \cite{Sanchez-Kuntz:2022gds}. This setup uses a 3-step scale-factor evolution, with two transition times $t_\mathrm{i} < t_\mathrm{f}$ and a measurement time $t_\mathrm{m} > t_\mathrm{f}$. The scale factor is schematically sketched in figure \ref{fig:scale_factor}:
\begin{enumerate}[(I)]
    \item $a(t) = a_\mathrm{i}$ for $t \leq t_\mathrm{i}$ with a constant initial scale factor $a_\mathrm{i}$, at which the experiment is prepared. We generally choose $a_\mathrm{i} = 1$ for simplicity.
    \item $a(t)$ for $t_\mathrm{i} \leq t \leq t_\mathrm{f}$ with a general time-dependent scale factor, for example the exponential or polynomial scale factors discussed before. $a(t)$ interpolates between $a_\mathrm{i}$ and $a_\mathrm{f}$ continuously.
    \item $a(t) = a_\mathrm{f}$ for $t \geq t_\mathrm{f}$ with a constant final scale factor $a_\mathrm{f}$ at which the measurement of the correlator is performed, which happens at time $t_\mathrm{m} > t_\mathrm{f}$.
\end{enumerate}
The mode functions for this setup are obtained by solving the EOM \eqref{eq:general_EOM} in each of the above phases and then defining the glued overall solution, demanding it to be continuous and continuously differentiable:
\begin{align}
    \phi_k(t) = \begin{cases}
        \phi^\RN{1}_k(t) & \mathrm{for} \ t < t_\mathrm{i} \\
        \phi^\RN{2}_k(t) & \mathrm{for} \ t_\mathrm{i} \leq t \leq t_\mathrm{f} \\
        \phi^\RN{3}_k(t) & \mathrm{for} \ t_\mathrm{f} < t
    \end{cases}.
\end{align}
The details of solving for the $\phi^{\RN{1} / \RN{2} / \RN{3}}_k(t)$ and gluing the solutions together at the transition times $t_\mathrm{i}$ and $t_\mathrm{f}$ are already discussed in \cite{Sanchez-Kuntz:2022gds} and can be found in appendix \ref{app:growing_modes}.

\subsubsection*{Integration}

\begin{figure}
    \centering
    \makebox[\textwidth][c]{\includegraphics{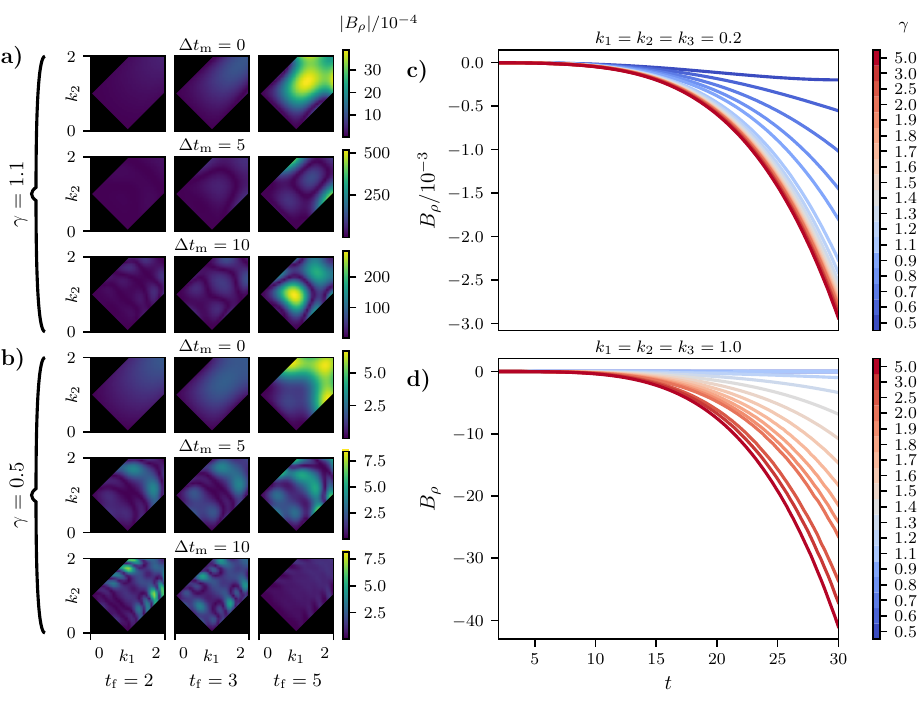}}
    \caption{Density bispectrum $B_\rho$. \textbf{a)} and \textbf{b)}: As a function of $k_1$ and $k_2$ at fixed $k_3 = 1.0$ and fixed $t_\mathrm{i} = 1$, for two different $\gamma$. We plot the absolute value which makes the patterns easier to read off, but both positive and negative values can be reached. Shown are multiple parameters in a $30 \times 30$ grid, while the $k_1$ and $k_2$ axes are shared. The triangles in the corners mark the kinematically excluded regions $k_a + k_b < k_c$ for $\{a, b, c\} = \{1, 2, 3\}$. The columns show different $t_\mathrm{f}$, while the rows in each block show different wait times $\Delta t_\mathrm{m} = t_\mathrm{m} - t_\mathrm{f}$, as indicated above each row. Larger $t_\mathrm{f}$ usually lead to larger values, and larger $\Delta t_\mathrm{m}$ lead to a interference-like redistribution of the values. $\gamma = 1.1$ leads to larger values than $\gamma = 0.5$. \textbf{c)} and \textbf{d)}: Time dependence at $k_1 = k_2 = k_3 = 0.2$ [a)] and $k_1 = k_2 = k_3 = 1.0$ [b)], at different $\gamma$ as a function of time $t = t_\mathrm{f} = t_\mathrm{m}$. All contributions grow at late times, hinting at IR divergences. Generally, larger $\gamma$ lead to faster late-time growth, and larger $k$ lead to much larger overall values. We have fitted all curves qualitatively with the predicted late-time dependence from table \ref{tab:late_time_3pt_functions} and the predicted dependencies match up to inaccuracies that can stem from numerical error and transient contributions from early times.}
    \label{fig:growing_plots_derivex}
\end{figure}

\begin{figure}[!htbp]
    \centering
    \makebox[\textwidth][c]{\includegraphics{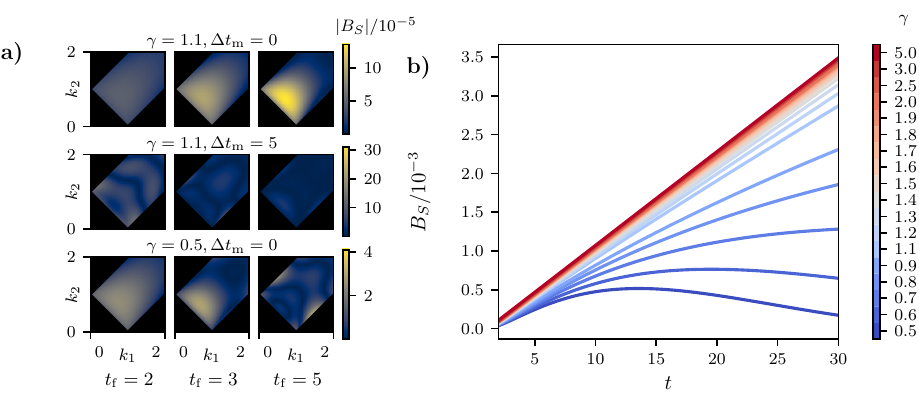}}
    \caption{Phase bispectrum $B_S$. The general structure of this figure is the same as for figure \ref{fig:growing_plots_derivex}. \textbf{a)}: As a function of $k_1$ and $k_2$ at fixed $k_3 = 1.0$. Similarly to $B_\rho$, the overall size of the values is growing with $t_\mathrm{f}$, while larger $\Delta t_\mathrm{m}$ lead to stronger patterns. Again, larger $\gamma$ lead to larger values overall. \textbf{b)}: Late-time behavior at $k_1 = k_2 = k_3 = 0.2$ at different $\gamma$ as a function of final time $t = t_\mathrm{f} = t_\mathrm{m}$. Values $\gamma < 1$ lead to oscillations and $\gamma > 1$ lead to growth, which matches the dependence on $t_\mathrm{f}$ from table \ref{tab:late_time_3pt_functions} qualitatively. Since the growth is linear and not quartic, larger values can be reached at earlier times than for $B_\rho$.}
    \label{fig:growing_plots}
\end{figure}

As explained in section \ref{sec:observables}, there are two kinds of observables in the BEC system: Direct correlation functions of the normalized density contrast $\delta\rho/\rho_0$ of the condensate, which translate to rescaled correlation functions of the derivative field $\dot\phi$, and homodyne measurements of the phase, which translate to to correlation functions of the field $\phi$ itself. We use the 3-point correlators, also called bispectra, defined earlier:
\begin{subequations}
\begin{align}
     B^3_\rho(k_1, k_2, k_3) &\coloneq -\kappa^3 \, a^6(t_\mathrm{m}) \ \langle \dot\phi_{k_1}(t_\mathrm{m}) \dot\phi_{k_2}(t_\mathrm{m}) \dot\phi_{k_3}(t_\mathrm{m}) \rangle^\prime, \label{eq:b_rho_def} \\
    B^3_S(k_1, k_2, k_3) &\coloneq \kappa^3 \, \langle \phi_{k_1}(t_\mathrm{m}) \phi_{k_2}(t_\mathrm{m}) \phi_{k_3}(t_\mathrm{m}) \rangle^\prime.
\end{align}
\end{subequations}
In the following, we will omit the superscript 3. Both observables have exactly one contribution at leading order from the 3 field interaction term that goes like $\kappa$. The bispectra can then be computed from the piece-wise mode functions as
\begin{subequations} \label{eq:B_definitions}
\begin{align}
    B_{\rho}(k_1, k_2, k_3) &= -\kappa^4 a_\mathrm{f}^6 \, \mathrm{Im} \Bigg[\Big(\dot\phi_{k_1}^\RN{3}(t_\mathrm{m})\dot\phi_{k_2}^\RN{3}(t_\mathrm{m})\dot\phi_{k_3}^\RN{3}(t_\mathrm{m})\Big)^* \notag\\
    & \qquad \times \Bigg( \int\limits_{-\infty^-}^{t_\mathrm{m}} \d t\, a^2(t) \dot\phi_{k_1}(t) \phi_{k_2}(t) \phi_{k_3}(t) \ \vec k_2 \cdot \vec k_3  + \mathrm{permutations}  \Bigg) \Bigg] \label{eq:B_rho} \\
    B_{S}(k_1, k_2, k_3) &= \kappa^4 \, \mathrm{Im} \Bigg[\Big(\phi_{k_1}^\RN{3}(t_\mathrm{m})\phi_{k_2}^\RN{3}(t_\mathrm{m})\phi_{k_3}^\RN{3}(t_\mathrm{m})\Big)^* \notag\\
    & \qquad \times \Bigg( \int\limits_{-\infty^-}^{t_\mathrm{m}} \d t\, a^2(t) \dot\phi_{k_1}(t) \phi_{k_2}(t) \phi_{k_3}(t) \ \vec k_2 \cdot \vec k_3  + \mathrm{permutations}  \Bigg) \Bigg] \label{eq:B_phi}
\end{align}
\end{subequations}
Using momentum conservation, the scalar products of the momenta can be expressed as
\begin{align}
    \vec k_1 \cdot \vec k_2 = \frac{k_3^2 - k_1^2 - k_2^2}{2} \quad \mathrm{and \ permutations}.
\end{align}
Since the scale factor is constant there, the first and third region of the integral yield an easy to calculate contribution. The complicated region is the second. Since the mode functions consist of Bessel functions, the above integrals cannot be evaluated analytically in their full generality. Therefore, we resort to numerical evaluation for some realistic numerical parameters. We use the \textsc{vegas} Monte-Carlo integration algorithm \cite{lepage_adaptive_2021}. The parameters that stay the same for all following numerical evaluations are $t_\mathrm{i} = 1$ and $k_{1/2} \in [0.01, 2.0]$. We use $10$ warm-up integration runs with $1000$ integrand evaluations and then $10$ integration runs with $10000$ evaluations. Results for $B_\rho$ are presented in figure \ref{fig:growing_plots_derivex} and results for $B_S$ are presented in figure \ref{fig:growing_plots}.

\subsection{Late-time Behavior and IR-Divergence}

A very interesting feature of the 3-point function is the late-time growth, as seen in figures \ref{fig:growing_plots_derivex} and \ref{fig:growing_plots}. Already in \cite{Weinberg:2005vy}, a theorem has been formulated that classifies which interaction vertices produce growing correlations at late times. However, this theorem implicitly assumes exponential growth of the scale-factor, as is typical in inflationary models. Therefore, it does not hold for other scale factors explored here, such as the polynomial $a(t) = Q t^\gamma$. Because of this, we shall repeat the analysis manually for the vertices in \eqref{eq:vertices}.

As shown in \cite{Weinberg:2005vy}, the $d$-dimensional mode equation can often be solved in terms of a power series in inverse powers of $a(t)$. For late times, we assume $k^2/a^2 \ll H^2 = \dot a^2 / a^2$, so that we can solve \eqref{eq:general_EOM} to zero-th order, ignoring the small third term, as
\begin{align}
    \phi^{(0)}_k = D_k + C_k \int\limits_t^\infty \frac{\d t'}{a^d(t')}.
\end{align}
With this, we return to the full \eqref{eq:general_EOM} and insert the zero-th order to solve self-consistently
\begin{align}
    \partial_t \Big( a^d \dot\phi \Big) &= k^2 a^{d-2} \phi \notag\\
    \Rightarrow \quad\quad \phi^{(1)}_k(t) &= D_k \left( 1 + k^2 \int\limits_{-\infty}^t \frac{\d t'}{a^{d}(t')} \int\limits_{-\infty}^{t'} \d t''\, a^{d-2}(t'') \right) \notag\\
    &\quad + C_k \left( \int\limits_t^\infty \frac{\d t'}{a^d(t')} + k^2 \int\limits_{-\infty}^t \frac{\d t'}{a^{d}(t')} \int\limits_{-\infty}^{t'} \d t''\, a^{d-2}(t'') \int\limits_{-\infty}^{t''} \frac{\d t'''}{a^d(t''')} \right) \notag\\
    &= C_k \psi_k(t) + D_k \mu_k(t), \label{eq:general_mode_function_expansion}
\end{align}
where, assuming a sufficiently fast-growing scale factor,
\begin{align}
    \psi_k(t) \xlongrightarrow{t\rightarrow\infty} 0, \qquad \mu_k(t) \xlongrightarrow{t\rightarrow\infty} 1.
\end{align}
This general expansion contains the scale factors $a(t) = e^{Ht}$ and $a(t) = Q \, t^\gamma$ for $\gamma > 1$ as a special case, as can be checked with \eqref{eq:de_sitter_massless_late_time} and \eqref{eq:polynomial_late_time} in appendix \ref{app:growing_modes}. In both cases, the first sub-leading correction comes from the second term in the first line, which is proportional to the leading order $D_k$. The functions $\psi_k(t)$ and $\mu_k(t)$ are real.

The decelerating polynomial scale factor with $\gamma < 1$ is not contained in the above generalization, since the physical momentum does not fall off quickly enough and the expansion \eqref{eq:general_mode_function_expansion} diverges. Still, we can split the exact expression \eqref{eq:decelerating_gamma_mode_function} into $C_k \psi_k(t)$ and $D_k \mu_k(t)$ and keep the same hierarchy of terms. The main difference is that both $\mu_k$ and $\psi_k$ vanish as $t \rightarrow \infty$, but also oscillate - there is no freeze-out.

The early-time dynamics are not important for the late-time properties. Therefore, at this stage, we ignore any matching to initial conditions and we focus entirely on the expansion phase in figure \ref{fig:scale_factor}. For simplicity, we cut off any time integrals arbitrarily at the lower bound and omit the earlier contributions. Lastly, we take $t_\mathrm{f} = t_\mathrm{m}$ for this discussion.

To study the late-time behavior, we can insert the split \eqref{eq:general_mode_function_expansion} into the expressions for $B_\rho$ and $B_S$ in \eqref{eq:B_definitions}. The terms involving $C_k \psi_k(t_\mathrm{m})$ or its time derivative are suppressed in comparison to the leading term with $D_k \mu_k(t_\mathrm{m})$ or its time derivative. However, since $\psi_k(t)$ and $\mu_k(t)$ are real functions for all $t$, the overall leading term is proportional to $| D_k |^2$ and therefore purely real. It cancels out due to taking the imaginary part. The next sub-leading terms are those that have exactly one $C_k$ term from either the final time mode functions evaluated at $t_\mathrm{m}$, or the vertex mode functions evaluated at $t$, which is integrated over. The leading terms are
\begin{subequations} \label{eq:leading_terms}
\begin{align}
    B^\mathcal{A}_{\rho / S} &\coloneq a_\mathrm{f}^{(6)} \, \mathrm{Im} \Bigg[ D_{k_1}^* C_{k_1} |D_{k_2}|^2 |D_{k_3}|^2 \, \overset{\scriptscriptstyle{(}\cdot\scriptscriptstyle{)}}{\mu}_{k_1}(t_\mathrm{m}) \overset{\scriptscriptstyle{(}\cdot\scriptscriptstyle{)}}{\mu}_{k_2}(t_\mathrm{m}) \overset{\scriptscriptstyle{(}\cdot\scriptscriptstyle{)}}{\mu}_{k_3}(t_\mathrm{m}) \notag\\
    & \qquad \qquad \qquad \times \Bigg( \int\limits_{-\infty^-}^{t_\mathrm{m}} \d t\, a^2(t) \dot\psi_{k_1}(t) \mu_{k_2}(t) \mu_{k_3}(t) + \psi_{k_1}(t) \dot\mu_{k_2}(t) \mu_{k_3}(t)  \Bigg) \Bigg] \\
    B^\mathcal{B}_{\rho / S} &\coloneq a_\mathrm{f}^{(6)} \, \mathrm{Im} \Bigg[ C_{k_1}^* D_{k_1} |D_{k_2}|^2 | D_{k_3}|^2 \,  \overset{\scriptscriptstyle{(}\cdot\scriptscriptstyle{)}}{\psi}_{k_1}(t_\mathrm{m}) \overset{\scriptscriptstyle{(}\cdot\scriptscriptstyle{)}}{\mu}_{k_2}(t_\mathrm{m}) \overset{\scriptscriptstyle{(}\cdot\scriptscriptstyle{)}}{\mu}_{k_3}(t_\mathrm{m}) \notag\\
    & \qquad \qquad \qquad \times \Bigg( \int\limits_{-\infty^-}^{t_\mathrm{m}} \d t\, a^2(t) \dot\psi_{k_1}(t) \mu_{k_2}(t) \mu_{k_3}(t) \Bigg) \Bigg]
\end{align}
\end{subequations}
We have omitted constants and most permutations, since we only care about the parametric late-time behavior. In the $B^\mathcal{A}$ term, we still allow the time derivative to hit either $\mu$ or $\psi$ to see which ever leads to a larger contribution. These expressions are for the $\kappa$ interaction in \eqref{eq:vertices}, we also define an equivalent term for the $\lambda$ interaction.

While the exponential and accelerating polynomial scale factors are straight-forward to integrate here, the decelerating polynomial ones are a bit more subtle because of the oscillation. Refer to appendix \ref{app:growing_modes} for a discussion of this integration.

The left half of table \ref{tab:late_time_3pt_functions} shows the late-time dependence of $B_\rho$ on $t_\mathrm{m}$, for both exponential and polynomial scale factors (accelerating and decelerating) and both interaction vertices featured in \eqref{eq:vertices}. From there, we can read off that, in fact, the vertices realized in the quantum simulator do produce IR divergences for a range of $\gamma$, and sometimes even independent of $\gamma$. This is also true for the phase correlation functions $B_S$, as can be seen in the right half of table \ref{tab:late_time_3pt_functions}.
\begin{table}
    \centering
    \makebox[\textwidth][c]{%
    \begin{tabular}{c||c|c|c||c|c|c|}
        & & $a^2(t) \lambda \phi^3 \vphantom{\Big(}$  & $a^2(t) \kappa \dot\phi k^2 \phi^2$ & & $a^2(t) \lambda_1 \phi^3 \vphantom{\Big(}$ & $a^2(t) \kappa \dot\phi k^2 \phi^2$ \\\hline\hline
        \multirow{2}{2cm}{exponential\\(de Sitter)} & $B_\rho^\mathcal{A}\vphantom{\Bigg(}$ & $\mathbf{t_\mathrm{m}^4}$ & $\mathbf{t_\mathrm{m}^4}$ & $B_S^\mathcal{A}\vphantom{\Bigg(}$ & $\mathbf{t_\mathrm{m}}$ & $\mathbf{t_\mathrm{m}}$ \\
                                     & $B_\rho^\mathcal{B}\vphantom{\Bigg(}$ & $\mathbf{t_\mathrm{m}^{2} e^{2Ht_\mathrm{m}}}$ & $\mathbf{t_\mathrm{m}^4}$ & $B_S^\mathcal{B}\vphantom{\Bigg(}$ & $1$ & $t_\mathrm{m}^2 \, e^{-2Ht_\mathrm{m}}$ \\\hline
        \multirow{2}{2cm}{accelerating\\polynomial\\with $\gamma>1$} & $B_\rho^{A}\vphantom{\Bigg(}$ & $\mathbf{t_\mathrm{m}^{5}}$ & $\mathbf{t_\mathrm{m}^{4}}$ & $B_S^\mathcal{A}\vphantom{\Bigg(}$ & $\mathbf{t_\mathrm{m}^2}$ & $\mathbf{t_\mathrm{m}^{1}}$ \\
                                     & $B_\rho^\mathcal{B}\vphantom{\Bigg(}$ & $\mathbf{t_\mathrm{m}^{3+2\gamma}}$ & $\mathbf{t_\mathrm{m}^4}$ & $B_S^\mathcal{B}\vphantom{\Bigg(}$ & $\mathbf{t_\mathrm{m}^2}$ & $\mathbf{t_\mathrm{m}^{3-2\gamma}}$ \\\hline
        \multirow{2}{2cm}{decelerating\\polynomial\\with $\gamma<1$} & $B_\rho^{A}\vphantom{\Bigg(}$ & $\mathbf{t_\mathrm{m}^{3\gamma}} \times e^{i f(t)}$ & $\mathbf{t_\mathrm{m}^{2\gamma}}e^{i f(t)}$ & $B_S^\mathcal{A}\vphantom{\Bigg(}$ & $1 \times e^{i f(t)}$ & $t_\mathrm{m}^{-\gamma} \times e^{i f(t)}$ \\
                                     & $B_\rho^\mathcal{B}\vphantom{\Bigg(}$ & $\mathbf{t_\mathrm{m}^{3\gamma}} \times e^{i f(t)}$ & $\mathbf{t_\mathrm{m}^{2\gamma}}e^{i f(t)}$ & $B_S^\mathcal{B}\vphantom{\Bigg(}$ & $1 \times e^{i f(t)}$ & $t_\mathrm{m}^{-\gamma} \times e^{i f(t)}$
    \end{tabular}}
    \caption{The leading late-time $t_\mathrm{m}$-dependence of the bispectra $B_\rho$ and $B_S$, as defined in \eqref{eq:leading_terms}. Shown are both terms that might be leading, $\mathcal{A}$ and $\mathcal{B}$, for all 3 considered scale factors and all interaction terms in \eqref{eq:vertices}. The left half displays $B_\rho$, the directly observable density bispectrum. The $e^{if(t)}$ for decelerating polynomials is supposed to indicate that the late-time behavior \textit{might} be oscillating. All interactions lead to divergencies for some combination of scale-factor and parameter $\gamma$, such that all observables have growing late-time contributions, which we call IR divergences. This strong trend is mostly supported by the $a^6(t)$ time dependence that is included in the observable \eqref{eq:b_rho_def}. The right hand side shows the homodyne observable $B_S$. There are no additional powers of the scale factor, so this observable is not as strongly divergent. Still, we expect late-time growth in the BEC correlator for exponential and accelerating polynomial scale factors.}
    \label{tab:late_time_3pt_functions}
\end{table}

\section{Periodic Scale Factors} \label{sec:periodic_a}

Unfortunately, it is not possible to manipulate the scale factor $a(t)$ arbitrarily for a realistic experiment. In the setup that we are focusing on, it may only be increased by a factor of around 3 before the assumptions of our effective theory start to break down. For this reason, it is difficult to expand the scale factor long enough for the expected growth in the correlator to take over as shown in figure \ref{fig:growing_plots_derivex}.

Instead, as a toy model for investigating growing correlation functions, we may turn to a different kind of driven system that has an oscillating scale factor with period $T$. Such systems have been previously considered in the context of pattern formation and reheating \cite{Liebster:2023mmf, Chatrchyan:2020cxs}. We assume that the scale factor never reaches $0$ and is bounded from above. We further parametrize the scalar field $\phi = \chi / a(t)$ to rephrase the EOM \eqref{eq:general_EOM}
\begin{align}
    \ddot \phi + d \frac{\dot a}{a} \dot\phi + \frac{k^2}{a^2} \phi = 0 \quad \Leftrightarrow \quad \ddot\chi + \underbrace{\left( \frac{k^2}{a^2(t)} - \frac{\ddot a(t)}{a(t)} \right)}_{\eqcolon \, V(t)} \chi = 0. \label{eq:hill_eq}
\end{align}
The EOM for $\chi$ is a so-called Hill equation. From Floquet theory, it is expected that solutions to this equation are superpositions of basis solutions of the form
\begin{align}
    \chi(t) = e^{F t} P(t), \label{eq:floquet_solution}
\end{align}
where $P(t)$ is a periodic function with period $T$, just like $V(t)$. These solutions are called Floquet solutions. The exponential prefactor $e^{F t}$ with general complex $F$ allows for different overall periods, as well as resonant growth or suppression.

\subsection{Resonant Particle Production}

\begin{figure}
    \centering
    \includegraphics{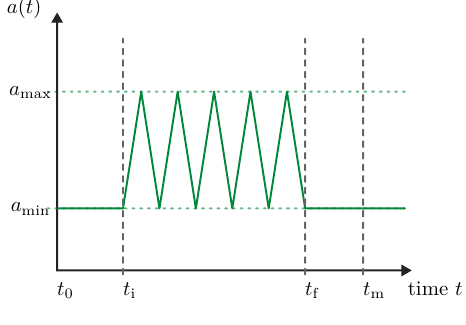}
    \caption{Periodic scale factor $a(t)$ as a function of time. The experiment is set up at time $t_0$ with a scale factor of $a(t_0) = a_\mathrm{min}$. Between the times $t_\mathrm{i}$ and $t_\mathrm{f}$, the scale factor is continuously oscillating between the values $a_\mathrm{min}$ and $a_\mathrm{max}$. The correlation function then is measured at time $t_\mathrm{m}$, when the scale factor has been held at a constant value $a_\mathrm{min}$ for some time.}
    \label{fig:periodic_scale_factor}
\end{figure}

For simplicity, we specify the scale factor to a simple triangle wave, as this case is solvable analytically. Figure \ref{fig:periodic_scale_factor} shows the experimental setup, it is analogous to the one we used in the growing scale factor discussion.

In principle, it is conceivable that the precise time dependence of $a(t)$ is important for particle production. In particular, table \ref{tab:late_time_3pt_functions} shows that the late-time behavior depends strongly on the scale factor. However, these effects are expected to only take over after the scale factor has expanded to a value of at least $a(t) = \O(10)$, as seen in figure \ref{fig:growing_plots_derivex}. The motivation for studying periodic systems in the first place was that we cannot access these large scale factors. We are shifting focus towards resonant interactions in this section. And these resonances should be much less sensitive to the precise form of the scale factor, as long as it is periodic and only varies over a small range. Therefore, we expect other wave-forms to not add anything qualitatively new.

\begin{figure}
    \centering
    \includegraphics{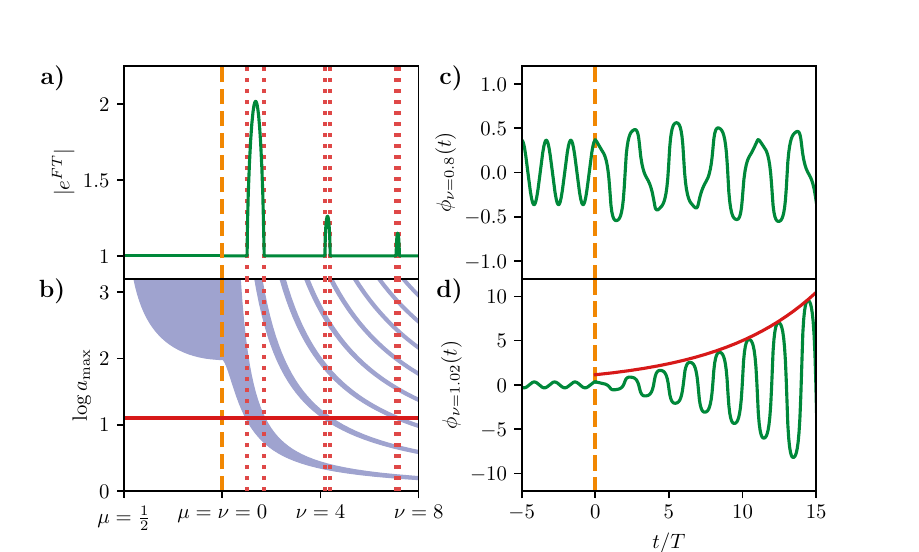}
    \caption{Resonant growth of the mode function and particle production. \textbf{a)} Absolute value of the growth factor $|e^{F T}|$ for the growing Floquet solution as a function of $k$, parametrized via $\mu$ and $\nu$, according to \eqref{eq:mudef} and \eqref{eq:nudef}. There are 3 bands in the displayed $k$-range where the value is larger than $1$. In these bands, exponential growth of the solution is expected. \textbf{b)} Visualization of the condition \eqref{eq:floquet_growth_condition}. Values of $\mu$ and $\nu$ for which $\log a_\mathrm{max}$ intersects the colored areas see exponential growth. We set $a_\mathrm{min} = 1$, so the horizontal line indicates $a_\mathrm{max} = 3$, and the dotted lines indicate the intersections. The dashed line indicates the transition from the low momentum to the high momentum regime. \textbf{c)} Sample off-resonant mode function. At time $t / T = 0$, indicated by the dashed line, the periodic modulation of the scale factor is turned on. The chosen $\nu = 0.8$ is below the first resonant regime and therefore, the mode function does not grow. \textbf{d)} Sample resonant wave function. Starting from time $t / T = 0$, the mode function exhibits exponential growth with the expected $\alpha$, as indicated by the red line.}
    \label{fig:4pictures}
\end{figure}

The details of deriving the mode function are discussed in appendix \ref{app:periodic_modes}. The main take-away is that there are infinitely many finite ranges of resonant $k$, parametrized via
\begin{align}
    \mu = \sqrt{\frac{1}{4} - k^2 \alpha^2} \quad \mathrm{or} \quad \nu = \sqrt{k^2 \alpha^2 - \frac{1}{4}},
\end{align}
whichever is real. Here, $\alpha$ is the slope of the scale factor in the linear intervals. These resonant bands are depicted in figure \ref{fig:4pictures}. The resonant $k$ lead to exponential growth of the mode function via \eqref{eq:floquet_solution}. Since the two-point function is simply the square of the mode function, this implies resonant growth of the two-point function. After some finite time, the EFT breaks down and a new mechanism results in a saturation of this particle production. The exact mechanism for this is beyond the scope of this work, but has been treated for example in \cite{Butera:2022kwi, Baak:2025ifl}. 

The final result is a strongly peaked 2-point function at the resonant $k$, which has been previously observed experimentally \cite{Liebster:2023mmf}. We claim that the features observed there are the result of the first resonant band seen in figure \ref{fig:4pictures}. However, because of the difference in exact time-dependence of the scale factor in that setup, the features can only be compared qualitatively.

\subsection{Resonant Interactions}

The main idea of this paper is studying interactions. To do this in the periodic scale factor model, we pose two conditions:
\begin{enumerate}
    \item The resonant two-point function growth should be switched off / suppressed
    \item The interactions should be resonant
\end{enumerate}
Together, these conditions guarantee that the interactions dominate the BEC dynamics and are not overshadowed by the 2-point function, like it would likely be the case when one would try to find interactions in the data of \cite{Liebster:2023mmf}. 

To illustrate the idea, consider the following toy model. Say that the mode functions behave approximately as $e^{ikt}$, except if they are resonant, where they grow exponentially. The scale factor goes approximately like (the real part of) $e^{i \omega t}$. Ignoring time derivatives for now, the 3-point function behaves like
\begin{align}
    \langle \phi_{k_1}(t) \phi_{k_2}(t) \phi_{k_3}(t) \rangle \propto \lambda \, \mathrm{Im} \left(e^{i (k_1 + k_2 + k_3) t} \int\limits_0^t\d t' a^n(t') e^{-i (k_1 + k_2 + k_3) t'}\right).
\end{align}
This integral becomes resonant if $n\omega - k_1 - k_2 - k_3 = 0$, where it grows linearly in time. If this interactive resonance condition can be met by lower $k$ than the the mode function resonance condition, then it might be possible to choose parameters such that the mode function resonant growth is not possible inside the validity range of the EFT, while the interaction still is resonant. Then, interactions should be straightforward to detect. Derivatives in the interaction should not qualitatively alter this picture, since they only produce factors of $\vec k$ or $|\vec k| / a$, which do not shrink at late times like in the growing scale factor scenario. 

\begin{figure}
    \centering
    \makebox[\textwidth][c]{\includegraphics{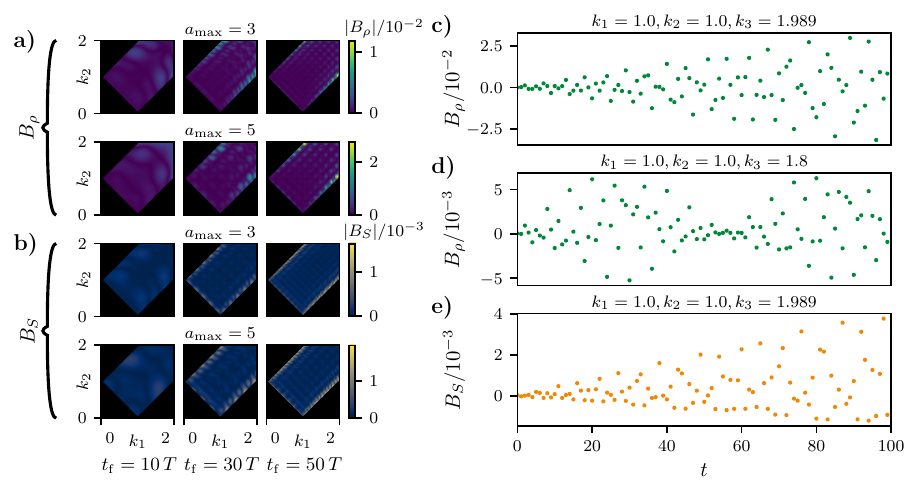}}
    \caption{Bispectra in the periodic scale factor system. \textbf{a)} and \textbf{b)} $B_\rho$ and $B_S$ as functions of $k_1$ and $k_2$ at fixed $k_3 = 1.0$ and fixed $T = 1$. Shown are $a_\mathrm{max} = 3$ and $a_\mathrm{max} = 5$, with fixed $a_\mathrm{min} = 1$ on a $50 \times 50$ grid. A wave-like structure with clear peaks around the boundaries of the kinematic region is visible. The effective wavelength goes down over time, while the height of the peaks increases. This general structure is the same for $B_\rho$ and $B_S$. The only difference is that because of the extra factors of $k$, values in the lower left corner of the picture are suppressed for $B_\rho$, but not for $B_S$. \textbf{c)} to \textbf{e)} Time series data for $B_\rho$ [c) and d)] and $B_S$ [e)]. The interaction can lead to resonant interactions [c)] that produce growing correlations $B_\rho$ over time, close to the kinematic boundary. For other values of $k$ [d)], the bispectrum is oscillating without growth. The same resonance / non-resonance happens for $B_S$ [e)].}
    \label{fig:3_pt_periodic}
\end{figure}

Returning to the full model, figure \ref{fig:3_pt_periodic} shows the analytic results of the evaluation of the $B_\rho$ and $B_S$ bispectra using the integral formula derived in appendix \ref{app:periodic_modes}. Empirically, there is a clear resonant band around $|k_1 \pm k_2 \pm k_3| = 0$, in which both bispectra seem to grow without bounds. Also depicted is the time dependence of the bispectra at resonant and non-resonant triples of $k$ values.  Due to the fluctuating coefficient in front of the integral, the time evolution seems to follow multiple trajectories, jumping between them after every period. For non-resonant $k$, this time evolution can sometimes even appear to be chaotic, but for resonant $k$, the envelope of the multiple trajectories grows linearly over time.

\section{Conclusions}

We have computed the first few interaction terms in the quantum simulator model, going past the free theory studied in \cite{Sanchez-Kuntz:2022gds}. For this, we have changed the parametrization of the fluctuations to a radial approach with phase and density fluctuations $\sigma$ and $\delta\rho$. The leading interaction terms are similar to those in the EFT of inflation model \cite{Cheung:2007st}. Therefore, BECs could be used to simulate phenomenologically relevant models of the early universe. This could be used to prepare theoretical and data-analysis related methods for a later measurement of the bispectrum in the actual cosmic microwave background or the large scale structure of the universe. 

We have examined the effects of these leading interaction term on the bispectra of the density and phase of the BEC, going through the quantum field $\phi$ and its time derivative. We have considered 2 scenarios: A cosmologically relevant growing scale factor, and a periodically oscillating scale factor.

The growing scale factor scenario produces a clear pattern in the 3-point function, see figures \ref{fig:growing_plots_derivex} and \ref{fig:growing_plots}. The overall size of the density bispectrum seems to be of order $10^{-4}$ to $10^{-2}$ in the realistic parameter range, depending on the exact parameters. Since we are talking about a repeatable experiment in the laboratory, it might be possible to detect these effects in the near future. However, it seems hard to develop shape functions for these patterns in the same way as is done in actual cosmological data analysis (see, e.g., \cite{Fergusson:2008ra}) because of the lack of symmetries and the large number of parameters ($\gamma$, $t_\mathrm{i}$, $t_\mathrm{f}$, $t_\mathrm{m}$). Therefore, more sophisticated statistical models might be necessary, which goes beyond the scope of this work.

A very interesting feature is the growth of the correlation functions over time for a very wide range of parameters, mainly for the more directly relevant density bispectrum, see table \ref{tab:late_time_3pt_functions}. The main caveat is that the growth only takes over after the scale factor has been increased by a number of order $10$ or larger, which is not experimentally feasible.

To circumvent this problem, we also consider periodically oscillating scale factors. By relying on resonance phenomena instead of late times, we do not need ever-growing scale factors, and produce arbitrarily large correlation functions under realistic conditions.

The main challenge here is that the 2-point function also experiences resonant growth in certain parameter regimes, as shown in figure \ref{fig:4pictures}. However, we propose fine-tuning the experimental parameters such that resonant 2-point functions involve very large momenta, effectively pushing them out of the validity range of the EFT. This would hopefully mean that the system is not dominated by resonant particle production and the 3-point correlator can be measured with sufficient signal-to-noise ratio. The question whether this is a viable strategy will have to be tested experimentally.

\subsection*{Limitations}

There are some limitations to our analysis. We have not studied the expected breakdown of the quantum simulator EFT that we expect for large correlation functions. This has been treated for example in \cite{Butera:2022kwi, Baak:2025ifl}, but these papers do not consider the impact on higher correlation functions. Probably, some variation of the stochastic inflation formalism \cite{Starobinsky:1986fx, Cespedes:2023aal} could be used to get more accurate answers. However, this is beyond the scope of this initial study. 

It is furthermore possible that other properties of the model system also have to be taken into account, for example boundary effects and heating effects from the driving of magnetic fields.

More fundamental limitations of our treatment are the small accessible scale factor range. This can likely not be resolved easily, since the density of the BEC has to be in a specific density range for condensation to happen at all. Since density scales with $a^2(t)$, this cannot be avoided and necessitates a completely different setup if one wants to study truly late-time effects like the ones shown in figure \ref{fig:growing_plots_derivex}. 

On the theoretical side, it seems that the Schwinger-Keldysh formalism is fundamentally constrained to only be able to compute averages. It is impossible to predict features of individual realizations like the positions and relative angles of the localized peaks in momentum space that make up the standing waves in the square lattice in \cite{Liebster:2023mmf}. A more direct treatment using the wave function of the condensate, like it was performed in that paper, seem more suited for these types of questions. Conceptually, this would go in the direction of field-level inference in cosmology (see, for example \cite{Leclercq:2025ywu}), which again goes beyond the scope of this paper.

\subsection*{Perspectives}

These limitations point us in some new directions. On the theory side, it is most natural to cross-check the results obtained here with a direct Gross-Pitaevski analysis. Alternatively, one could try to formulate a stochastic formalism, like mentioned above. 

To generalize the treatment presented here, a useful step would likely be to allow for position- and / or time-dependent background fields $\rho_0(t, \vec x)$ in the derivation of the action. This would facilitate a more accurate treatment in the periodically driven scale factor setup, as the emergent lattice configuration can be treated as part of the background that influences the higher correlation functions, rather than an obstruction to their direct detection. However, also this has limited use for cosmological systems, where we always assume homogeneity.

In order to circumvent some part of the constraints on the usable range of scale factors, it might be possible to specialize to a different experimental setup altogether, like one where the BEC is physically expanding \cite{PhysRevX.8.021021}. However, it is not clear if this would actually be able to evade the condensation bound mentioned above.

When it comes to the periodic scale factors, it might still be of value to perform a systematic scan of different wave-forms in order to find one that leads to the strongest resonant divergences. It seems likely that a pure sine wave or some other analytic scale factor would perform better than the non-differentiable triangle wave. However, such a treatment most likely has to rely entirely on numerical studies.

Further, the fact that the BEC is coupled to the rest of the experimental setup, if only weakly, could be used as a test of the open EFT paradigm recently developed \cite{Salcedo:2024smn}.

As a last outlook, we want to mention that if one were to assume a cosmological model where the driving force of inflation is a coupling $\lambda(t)$, effectively running in time, possibly controlled via a symmetry-breaking background field, parts of our analysis in this paper can be directly translated into an exact cosmological model, so the applications might not be limited to BEC systems.

\acknowledgments

We thank Stefan Flörchinger and Markus Oberthaler for sharing their insights and for clarifications about the theoretical and experimental setup that we used. MN also would like to thank Ka Hei Choi, Carlo Cremonini, Avedis Neehus and Nick von Selzam for discussions. This work is supported in part  by the  German Excellence Strategy EXC 2094/2: ORIGINS 2. 
After a draft was prepared, we have used Anthropics Claude Fable 5 AI model to proofread and acknowledge that it found some issues that we subsequently corrected with its assistance.

\appendix
\section{EFT of Inflation} \label{app:dSEFT}

In this appendix, we want to briefly present the EFT of inflation, the effective field theory of gravitons interacting with a scalar field in a inflationary FLRW background, as first described in \cite{Cheung:2007st}. We do this to compare the Lagrangian of the interacting BEC \eqref{eq:scale_free_action_final} with a realistic cosmological Lagrangian. We specialize the arguments to 3 space-time dimensions. As argued in \cite{Cheung:2007st}, the most general Lagrangian for this system in unitary gauge, i.e.\ where the fluctuations of the massive scalar have been gauged away, is
\begin{align}
    S = \int \d^3x &\sqrt{-g} \Bigg( \frac{1}{2} \Mpl R - c(t) g^{00} - \Lambda(t) \notag\\
    &+ \frac{1}{2!} M_2^3(t) \big( g^{00} + 1 \big)^2 + \frac{1}{3!} M_3^3(t) \big( 1+ g^{00} \big)^3 + \frac{1}{4!} M_4^3(t) \big( 1+ g^{00} \big)^4 + \cdots  \Bigg).
\end{align}
We have omitted terms that involve the external curvature, since they are higher order in the derivatives. We have used $\Mpl = 1/(16\pi G_3)$ with the 3-dimensional gravitational constant $G_3$. Assuming a FLRW background metric
\begin{align}
    \d s^2 &= -\d t^2 + a^2(t) \d \vec x^2,
\end{align}
with $H = \dot a/ a$, the equations of motion, i.e.\ the modified Friedmann equations, are
\begin{subequations}
\begin{align}
    H^2 &= \frac{1}{\Mpl} \Big( c(t) + \Lambda(t) \Big), \\
    \dot H + H^2 &= \frac{1}{\Mpl} \Big( - c(t) + \Lambda(t) \Big), \\
    \partial_0 \Big( \dot H + H^2 \Big) &= 0.
\end{align}
\end{subequations}
The last equation represents energy conservation. From this, we can conclude
\begin{subequations}
\begin{align}
    c(t) &= - \frac{\Mpl}{2} \dot H, \\
    \Lambda(t) &= \Mpl \Big( H^2 + \frac{1}{2} \dot H \Big).
\end{align}
\end{subequations}
Note that these differ from the results found for the 4-spacetime-dimensional analysis in \cite{Cheung:2007st}. We can now perform the Stückelberg trick
\begin{align}
    g^{00} \longrightarrow \big( 1 + \dot\pi \big)^2 g^{00} + 2\big( 1 + \dot\pi \big) \partial_j \pi g^{0j} + \partial_j \pi \partial_k \pi g^{jk}.
\end{align}
Afterwards, we use the arguments in \cite{Cheung:2007st} to argue that in the appropriate energy range, mixing terms with the graviton $\delta g^{\mu\nu}$ are suppressed and the Goldstone boson $\pi$ decouples from the graviton.
We use canonical normalization of $\pi$, i.e.\ $\pi_\mathrm{c} = \sqrt{\Mpl \dot H} \pi$ and ignore the background graviton action. When computing $\dot\pi_\mathrm{c}$, we neglect the contribution proportional to $\ddot H$, since it is slow-roll suppressed. Up to 4-$\pi$ interaction terms, we find
\begin{align}
    S_\pi &= \int \d^3 x \, a^2(t) \Bigg[ \frac{1}{2} \partial_\mu \pi_\mathrm{c} \partial^\mu \pi_\mathrm{c} \notag\\
    &+ \frac{1}{2!} \frac{M_2^3(t)}{\Mpl^2 \dot H^2} \Bigg( 4\Mpl \dot H \dot\pi^2 + \sqrt{\Mpl \dot H} \Big( 4 \dot\pi^3 - \frac{4}{a^2(t)} \dot\pi (\partial_j \pi)^2 \Big) + \frac{1}{a^2(t)}\Big( -2\dot\pi^2(\partial_j \pi)^2 + \frac{1}{a^2(t)} (\partial_j \pi)^4 \Big) \Bigg) \notag\\
    &+ \frac{1}{3!} \frac{M_3^3(t)}{\Mpl^2 \dot H^2} \Bigg( -8 \sqrt{\Mpl \dot H} \dot\pi^3 -6\dot\pi^4 + \frac{6}{a^2(t)} \dot\pi^2 (\partial_j \pi)^2 \Bigg) \notag\\
    &+ \frac{1}{4!} \frac{M_4^3(t)}{\Mpl^2 \dot H^2} 16\dot\pi^4 + \cdots \Bigg]. \label{eq:dSEFT_action_final}
\end{align}

\section{Growing Scale Factor Mode Functions} \label{app:growing_modes}

In this appendix, we present the lengthy technical derivations of the mode functions $\phi_k(t)$ using the three different expanding scale factors laid out in the main text. Remember that we use the dimensionless quantities, rescaled by the natural scales of the problem.

\subsection{Solving the EOM}

First, we solve the mode EOM \eqref{eq:general_EOM} for the 3 highlighted scale factors: exponential, accelerating and decelerating polynomial.

\subsubsection*{Exponential Scale Factor}

We start with the de Sitter scale factor $a(t) = e^{H t}$. The EOM \eqref{eq:general_EOM} specializes to
\begin{align}
    \ddot\phi + d \, H \dot\phi + \frac{k^2}{e^{2Ht}} \phi = 0. \label{eq:eom_deSitter}
\end{align}
The solutions are
\begin{align}
    \phi_k(t) &= e^{-\frac{dHt}{2}} \Bigg( A \cdot H_1^{(1)}\Big( \frac{k}{H} e^{-Ht} \Big) + B \cdot H_1^{(2)}\Big( \frac{k}{H} e^{-Ht} \Big) \Bigg), \label{eq:general_deSitter_mode}
\end{align}
with constants $A$ and $B$ that are fixed by the initial conditions and the normalization. This will be done later on with the glueing of the individual phases of the experiment, depicted in figure \ref{fig:scale_factor}. 
For now, we consider late times, where the Hankel functions behave like
\begin{align}
    H_1^{(1)}(z) &= -\frac{2i}{\pi z} + \frac{i}{2\pi} \Bigg( 2\gamma_\mathrm{E} - 1 - i\pi + 2 \log\Big(\frac{z}{2}\Big) \Bigg) z + \O(z^2). \label{eq:H1_expansion}
\end{align}
The first term with $1/z$ is leading and the second term that goes like $z$ is subleading. Specializing to $d=2$, we can approximate for late times
\begin{align}
    \phi_k(t)
    \approx C_k \underbrace{e^{-2Ht}}_{\eqcolon \psi(t)} + D_k \underbrace{\Big( 1 + \# e^{-2H t} \big( \mathrm{const.} + H t \big) \Big)}_{\eqcolon \mu(t)}. \label{eq:de_sitter_massless_late_time}
\end{align}
This is a special case of the general \eqref{eq:general_mode_function_expansion}, for constants $C_k$ and $D_k$ that can be read off from the expansion of the Hankel function.

\subsubsection*{Polynomial Scale Factor}

The second scale factor we consider is $a(t) = Q \, t^\gamma$. The general EOM \eqref{eq:general_EOM} now reads
\begin{align}
    \ddot\phi + \frac{d\gamma}{t} \dot\phi + \frac{k^2}{Q^2 t^{2\gamma}} \phi &= 0,
\end{align}
which has solutions
\begin{align}
    \phi_k &= t^{\frac{1-d\gamma}{2}} \Bigg( A \ Y_{\frac{d\gamma-1}{2\gamma-2}} \Big( \frac{k}{Q} \frac{t^{1-\gamma}}{1-\gamma} \Big) + B \ J_{\frac{d\gamma-1}{2\gamma-2}}\Big( \frac{k}{Q} \frac{t^{1-\gamma}}{1-\gamma}\Big) \Bigg). \label{eq:polynomial_general_solution}
\end{align}
Due to the argument of the Bessel function, which scales as $t^{1-\gamma}$, we have to split the following discussion into the accelerating case $\gamma > 1$ and the decelerating case $\gamma < 1$. This is because the argument is $\frac{k t}{a(t)}$.
For $\gamma > 1$, this combination falls off, which corresponds to the physical wavelength growing faster than the Hubble horizon and a freeze-out of the modes. For $\gamma < 1$, the above combination grows, which corresponds to the Hubble horizon growing faster than the physical wavelength and, therefore, oscillations continue. This has already been discussed in \cite{Sanchez-Kuntz:2022gds}.

\subsubsection*{Accelerating Case $\gamma > 1$}
We use the power series representation
\begin{subequations}
\begin{align}
    J_\mu(z) &= \Big(\frac{z}{2}\Big)^\mu \sum\limits_{k=0}^\infty \frac{(-1)^k}{\Gamma(1+k+\mu) k!} \Big( \frac{z}{2} \Big)^{2k}, \\
    Y_\mu(z) &= \csc(\pi\mu) \Bigg[ \cos(\pi\mu) \Big( \frac{z}{2} \Big)^{\mu}  \sum\limits_{k=0}^\infty \frac{(-1)^k}{\Gamma(1+k+\mu) k!} \Big( \frac{z}{2} \Big)^{2k} - \Big( \frac{z}{2} \Big)^{-\mu} \sum\limits_{k=0}^\infty \frac{(-1)^k}{\Gamma(1+k-\mu) k!} \Big( \frac{z}{2} \Big)^{2k} \Bigg]
\end{align}
\end{subequations}
where we defined $\mu = \frac{d\gamma-1}{2\gamma-2}$. Inserting into \eqref{eq:polynomial_general_solution}, we get for late times
\begin{align}
    \phi_k &\approx C_k \underbrace{t^{1-d\gamma}}_{\eqcolon \psi(t)} + D_k \underbrace{\Big( 1 + \# t^{2-2\gamma} \Big)}_{\eqcolon \mu(t)}, \label{eq:polynomial_late_time}
\end{align}
with some factors $C_k$ and $D_k$ that can again be read off from the expansion.
Note that the term proportional to $D_k$ in \eqref{eq:polynomial_late_time} contains a constant term for late times, but also infinitely many terms that are barely sub-leading if $\gamma$ is only slightly bigger than $1$. However, since the infinite series only contains real coefficients and the only complex phase is contained in $C_k$ and $D_k$, these sub-leading terms will not contribute to the leading order late time behavior of the bispectrum. That is because terms that only contain $D_k$ or $C_k$ will cancel out after taking the imaginary part due to the Schwinger-Keldysh diagrammatical rules.

\subsubsection*{Decelerating Case $\gamma < 1$}

For $\gamma < 1$, we have to expand the Bessel functions not around argument $z \approx 0$, but rather large positive arguments $z \rightarrow + \infty$. We use the asymptotic expansion
\begin{subequations}
\begin{align}
    J_\mu(z) = \sqrt{\frac{2}{\pi z}} \Bigg( \cos\Big( z - \frac{\mu \pi}{2} - \frac{\pi}{4} \Big) + \frac{1-4\mu^2}{8} \frac{1}{z} \sin\Big( z - \frac{\mu\pi}{2} - \frac{\pi}{4} \Big) + \O\Big(z^{-2}\Big) \Bigg), \\
    Y_\mu(z) = \sqrt{\frac{2}{\pi z}} \Bigg( \sin\Big( z - \frac{\mu \pi}{2} - \frac{\pi}{4} \Big) - \frac{1-4\mu^2}{8} \frac{1}{z} \cos\Big( z - \frac{\mu\pi}{2} - \frac{\pi}{4} \Big) + \O\Big(z^{-2}\Big) \Bigg).
\end{align}
\end{subequations}
This time, inserting the expansions into \eqref{eq:polynomial_general_solution}, gives us for late times
\begin{align}
    \phi_k(t) &\approx C_k \ \underbrace{t^{\frac{\gamma}{2} - 1} \Bigg[ B \cos\Bigg( \overbrace{\frac{\frac{k}{Q} t^{1-\gamma} + \frac{\pi\gamma}{4}}{1-\gamma}}^{\eqcolon f(t)} \Bigg) + A \sin\Big( f(t) \Big) \Bigg]}_{\eqcolon \psi(t)}  \notag\\
    & \quad + D_k \  \underbrace{t^{-\frac{\gamma}{2}} \Bigg[ A \cos\Big( f(t) \Big) - B \sin\Big( f(t) \Big) \Bigg]}_{\eqcolon \mu(t)} \label{eq:decelerating_gamma_mode_function}
\end{align}
This expression is oscillating for late times, as expected, and the amplitude drops off to $0$ regardless of the initial conditions. The second term proportional to $D_k$ falls off less quickly than the other one and can therefore be considered to be leading order.

\subsection{Gluing of Mode Functions}

With the general mode functions \eqref{eq:general_deSitter_mode} and \eqref{eq:polynomial_general_solution}, we can now compute the full mode function over all three phases of expansion in the experimental setup depicted in figure \ref{fig:scale_factor}. This is for the numerical evaluation used in figures \ref{fig:growing_plots_derivex} and \ref{fig:growing_plots}, so we focus on the polynomial time dependence in the expanding time interval. The following discussion is a slight generalization of the appendices in \cite{Sanchez-Kuntz:2022gds}.

\subsubsection*{Early Times}

At early times, the space-time is Minkowski, since the expansion has not started yet. This gives us the typical EOM for a free field with the plane wave solution
\begin{align}
    \phi_k^\RN{1}(t) = A_k^\RN{1} e^{-i \frac{k t}{a_\mathrm{i}}} + B_k^\RN{1} e^{+i \frac{k t}{a_\mathrm{i}}} \label{eq:incoming_wave}
\end{align}
As usual, we set $B_k^\RN{1} = 0$ which corresponds to an incoming plane wave solution and then the Wronskian condition tells us
\begin{align}
    i &\overset{!}{=} a^d(t) \Big( \phi_k^\RN{1}(t) \big(\dot\phi_k^\RN{1}(t)\big)^* - \dot\phi_k^\RN{1}(t) \big(\phi_k^\RN{1}(t))^* \Big) \notag\\
    \Rightarrow \quad |A_k^\RN{1}|^2 &= \frac{1}{2 a_\mathrm{i} k}.
\end{align}

\subsubsection*{Intermediate Times with Polynomial Scale Factor}

We may use the solution of the EOM \eqref{eq:polynomial_general_solution}, which is, in new notation,
\begin{align}
    \phi_k^\RN{2}(t) &= t^{\frac{1}{2} - \gamma} \Bigg( A_k^\RN{2} Y_\frac{2\gamma-1}{2\gamma-2}\Big( \frac{k}{Q} \frac{t^{1-\gamma}}{1-\gamma} \Big) + B_k^\RN{2} J_\frac{2\gamma-1}{2\gamma-2}\Big( \frac{k}{Q} \frac{t^{1-\gamma}}{1-\gamma} \Big) \Bigg).
\end{align}
The overall mode function has to be continuous and differentiable, so we have to implement gluing conditions at $t = t_\mathrm{i}$ and infer the coefficients $A^\RN{2}_k$ and $B^\RN{2}_k$:
\begin{align}
    \phi_k^\RN{1}(t_\mathrm{i}) &\overset{!}{=} \phi_k^\RN{2}(t_\mathrm{i}) \quad \mathrm{and} \quad \dot\phi_k^\RN{1}(t_\mathrm{i}) \overset{!}{=} \dot\phi_k^\RN{2}(t_\mathrm{i}) \notag\\
    \Leftrightarrow \quad \begin{pmatrix}
        A_k^\RN{2} \\
        B_k^\RN{2}
    \end{pmatrix}
    &= - \frac{\pi}{2\sqrt{2}} t_\mathrm{i}^{\frac{1}{2} + \gamma} \frac{\sqrt{k} \, e^{\frac{-i k t_\mathrm{i}}{a_\mathrm{i}}}}{(1-\gamma) Q^{\frac{3}{2}}} \begin{pmatrix}
        J_\mathrm{i}' & - J_\mathrm{i} \\
        -Y_\mathrm{i}' & Y_\mathrm{i}
    \end{pmatrix} 
    \begin{pmatrix}
        1 \\
        -i
    \end{pmatrix},
\end{align}
where we defined the abbreviations
\begin{subequations}
\begin{alignat}{2}
    J_\mathrm{i} &= J_\frac{2\gamma-1}{2\gamma-2}\Big( \frac{k}{Q} \frac{t_\mathrm{i}^{1-\gamma}}{1-\gamma}\Big), \quad Y_\mathrm{i} &= Y_\frac{2\gamma-1}{2\gamma-2}\Big( \frac{k}{Q} \frac{t_\mathrm{i}^{1-\gamma}}{1-\gamma}\Big), \\
    J_\mathrm{i}' &= J_\frac{1}{2\gamma-2}\Big( \frac{k}{Q} \frac{t_\mathrm{i}^{1-\gamma}}{1-\gamma}\Big), \quad Y_\mathrm{i}' &= Y_\frac{1}{2\gamma-2}\Big( \frac{k}{Q} \frac{t_\mathrm{i}^{1-\gamma}}{1-\gamma}\Big).
\end{alignat}
\end{subequations}
Note that these fulfill a Wronskian condition themselves, namely
\begin{align}
    Y_\mathrm{i} J_\mathrm{i}' - J_\mathrm{i} Y_\mathrm{i}' = - \frac{2}{\pi} \frac{Q}{k} \frac{1-\gamma}{t_\mathrm{i}^{1-\gamma}}.
\end{align}

\subsubsection*{Late Times}

For late times, the constant scale factor again gives us plane wave solutions
\begin{align}
    \phi_k^\RN{3}(t) = A_k^\RN{3} e^{-i \frac{k t}{a_\mathrm{f}}} + B_k^\RN{3} e^{i \frac{k t}{a_\mathrm{f}}}.
\end{align}
We implement the same gluing condition for the second transition point at time $t = t_\mathrm{f}$. Defining $J_\mathrm{f}$ and $Y_\mathrm{f}$ like above, but for the final time, we find
\begin{align}
    \phi_k^{\RN{3}}(t_\mathrm{f}) &\overset{!}{=} \phi_k^\RN{2}(t_\mathrm{f}) \quad \mathrm{and} \quad \dot\phi_k^\RN{3}(t_\mathrm{f}) \overset{!}{=} \dot\phi_k^\RN{2}(t_\mathrm{f}) \notag\\
    \Leftrightarrow \quad \begin{pmatrix}
        A_k^\RN{3} \\
        B_k^\RN{3}
    \end{pmatrix} &= \frac{1}{2} t_\mathrm{f}^{\frac{1}{2}-\gamma} \begin{pmatrix}
        e^{i \frac{k t_\mathrm{f}}{a_\mathrm{f}}} Y_\mathrm{f} + i e^{i \frac{k t_\mathrm{f}}{a_\mathrm{f}}} Y_\mathrm{f}' & e^{i \frac{k t_\mathrm{f}}{a_\mathrm{f}}} J_\mathrm{f} + i e^{i \frac{k t_\mathrm{f}}{a_\mathrm{f}}} J_\mathrm{f}' \\
        e^{-i \frac{k t_\mathrm{f}}{a_\mathrm{f}}} Y_\mathrm{f} - i e^{-i \frac{k t_\mathrm{f}}{a_\mathrm{f}}} Y_\mathrm{f}' \qquad & e^{-i \frac{k t_\mathrm{f}}{a_\mathrm{f}}} J_\mathrm{f} - i e^{-i \frac{k t_\mathrm{f}}{a_\mathrm{f}}} J_\mathrm{f}'
    \end{pmatrix}
    \begin{pmatrix}
        A_k^\RN{2} \\
        B_k^\RN{2}
    \end{pmatrix}.
\end{align}

\subsection{Late-Time Bispectra}

The last computation that we have to elaborate on is the filling of table \ref{tab:late_time_3pt_functions}. For that, we need the late-time limits of all mode functions and their time derivatives. Note that the time derivatives fall off more quickly than the modes themselves. This leads to less strong late-time dynamics. The relevant proportionalities are
\begin{subequations}
\begin{alignat}{4}
    \dot\mu_k(t) &\sim t \, e^{-dHt} & &  & &\mathrm{exponential \ } & &a(t), \vphantom{\Big(} \\
    &\sim t^{1-d\gamma} & & \qquad\gamma > 1 \ & &\mathrm{polynomial \ } & &a(t), \vphantom{\Big(} \\
    &\sim t^{-\frac{3}{2}\gamma} e^{i f t^{1-\gamma}} & &\qquad\gamma < 1 \ & &\mathrm{polynomial \ } & &a(t); \vphantom{\Big(} \\
    \dot\psi_k(t) &\sim e^{-dHt} & &  & &\mathrm{exponential \ } & &a(t), \vphantom{\Big(} \\
    &\sim t^{-d\gamma} & &\qquad\gamma > 1 \ & &\mathrm{polynomial \ } & &a(t), \vphantom{\Big(} \\
    &\sim t^{-\frac{\gamma}{2} - 1} e^{i f t^{1-\gamma}} & &\qquad\gamma < 1 \ & &\mathrm{polynomial \ } & &a(t). \vphantom{\Big(}
\end{alignat}
\label{eq:expanding_mode_time_derivative}
\end{subequations}
To compute the entries of table \ref{tab:late_time_3pt_functions}, we haved use the late-time dependencies \eqref{eq:de_sitter_massless_late_time}, \eqref{eq:polynomial_late_time}, \eqref{eq:decelerating_gamma_mode_function} and \eqref{eq:expanding_mode_time_derivative} and inserted them into \eqref{eq:leading_terms}. While computing the subsequent integrals is straightforward for polynomials, it is a bit more tricky for the oscillating functions found for $\gamma < 1$. To compute their late-time bispectra, we write the oscillating part of the mode function as $e^{i f t^{1-\gamma}}$. This is always allowed as we can combine sines and cosines into some superposition of complex phases. When computing the three-point function, we always have 3 mode functions under the integral, which is an odd number. Therefore, the oscillation with positive and negative $f$ can never completely cancel and this template of an oscillating function stays valid. We therefore always integrate functions like
\begin{align}
    \int\limits^t \d t \, t^\alpha e^{i f t^{1-\gamma}} &\sim t^{1+\alpha} E_{\frac{\alpha + \gamma}{\gamma - 1}} \Big( -i f t^{1-\gamma} \Big) \notag\\
    &\sim t^{\alpha + \gamma} e^{i f t^{1-\gamma}}.
\end{align}
This involves the exponential integral function $E_{\nu}(z)$, which at late times goes like
\begin{align}
    E_\nu(z) \sim \frac{e^{-z}}{z}.
\end{align}
We can conclude the following: Integrating a polynomial in $t$ over time adds one power of $t$. Instead, integrating a polynomial times an oscillatory bit $e^{i f t^{1-\gamma}}$ inserts a factor of $t^\gamma$ as far as the parametric scaling is concerned.
Further, taking a derivative of a polynomial in $t$ removes one power of $t$. Taking a derivative of a polynomial times an oscillatory bit spawns two terms: One where the derivative acts on the polynomial and one where it acts on the phase. Acting on the phase gives a factor of $t^{-\gamma}$ as the interior derivative. This falls off slower for $\gamma < 1$ and therefore is the new dominant term. In summary: The derivative removes one factor of $t^\gamma$, as expected from the previous point.

Only after computing the integral, the external mode functions that come from the Schwinger-Keldysh rules might cancel the osciallatory phase to yield a non-oscillating term. This is why we noted in table \ref{tab:late_time_3pt_functions} that the late-time bispectra \textit{might} be oscillating. Numerically, we can see that they have oscillating components, as depicted in figure \ref{fig:growing_plots}. However, to obtain the qualitative fits mentioned there, we also allowed non-oscillating pieces, which was necessary to get the correct behavior.

\section{Periodic Scale Factor} \label{app:periodic_modes}

In this appendix, we present some details about the derivation of the mode functions in the periodic scale factor case. We have kept this treatment in a way that generalizes easily to any piecewise-defined scale factor for which the EOM \eqref{eq:general_EOM} can be solved analytically.

\subsection{Mode Functions}

In general, the Hill equation \eqref{eq:hill_eq} is very difficult to solve. Therefore, we focus on a scale factor $a(t)$ that is a triangle wave, oscillating between $a_\mathrm{min}$ and $a_\mathrm{max}$ using linear interpolation
\begin{align}
    a(t) = \begin{cases}
        a_\mathrm{min} + \alpha ( t - n T) & t \in \Big( nT, \big(n+\frac{1}{2}\big) T\Big) \\
        a_\mathrm{min} - \alpha ( t - (n+1) T) & t \in \Big( \big(n+\frac{1}{2}\big), (n+1) T\Big)
    \end{cases}
\end{align}
Here, $\alpha = 2(a_\mathrm{max} - a_\mathrm{min})/T$ is the slope of the scale factor in time. Then, $\ddot a(t) = 0$ everywhere but at the kinks, where the solution has to be glued together by demanding continuity of the mode function and a jump in the derivative given by
\begin{subequations}\label{eq:glue_conditions}
\begin{align}
    \lim_{\epsilon \rightarrow 0} \Big( \chi(t-\epsilon) - \chi(t+\epsilon) \Big) &= 0, \\
    \dot\chi(t + \epsilon) - \dot\chi(t - \epsilon) &= \int\limits_{t - \epsilon}^{t + \epsilon} \d t' \, \frac{\ddot a(t')}{a(t')} \chi(t'),
\end{align}
\end{subequations}
where the second time derivative of $a(t)$ is a comb of delta functions at the kinks
\begin{align}
    \ddot a(t) = \sum\limits_{n \in \mathbb{Z}} 2 \alpha \left( \delta(t-nT) - \delta\Big(t - \big(n+\frac{1}{2}\big)T\Big) \right).
\end{align}
For the solution of the EOM, there are now 3 regimes:
\begin{enumerate}
    \item long wavelength $k < \frac{\alpha}{2}$
    \item critical wavelength $k = \frac{\alpha}{2}$
    \item short wavelength $k > \frac{\alpha}{2}$
\end{enumerate}
The critical wavelength can be seen as the limit of either regime, so we will not discuss it further.

Next, the triangle wave of the scale factor has two phases: Expansion and contraction. We solve the EOM in each phase and then glue them together using \eqref{eq:glue_conditions}. Then, the next expansion phase is glued to the end of the previous contraction phase. Iteratively, the entire mode function can be obtained in this way. Since the problem is periodic, we define the periodicity-removed time coordinate
\begin{align}
    \tau = t - n T,
\end{align}
where $n \in \mathrm{Z}$ is chosen such that $\tau \in [0, T)$.

\subsubsection*{Long Wavelength}

In the long wavelength regime, the solutions for the mode functions are
\begin{align}
    \chi_\mathrm{ex}^\mathrm{lo}(\tau) = A_1 (a_\mathrm{min} + \alpha \tau)^{\frac{1}{2} + \mu} + B_1 (a_\mathrm{min} + \alpha \tau)^{\frac{1}{2} - \mu} \label{eq:low_k_expanding_mode}
\end{align}
in the expanding phase and
\begin{align}
    \chi_\mathrm{co}^\mathrm{lo}(\tau) = A_2 \Big(a_\mathrm{min} + \alpha (T-\tau)\Big)^{\frac{1}{2} + \mu} + B_2 \Big(a_\mathrm{min} + \alpha (T-\tau)\Big)^{\frac{1}{2} - \mu}
\end{align}
in the contracting phase. The exponent $\mu$ is defined as
\begin{align}
    \mu = \sqrt{\frac{1}{4} - \frac{k^2}{\alpha^2}}, \label{eq:mudef}
\end{align}
and the parameters $A_{1/2}$ and $B_{1/2}$ are determined by the initial and gluing conditions. The total solution then is
\begin{align}
    \chi^\mathrm{lo}(t) = \begin{cases}
        \chi^\mathrm{lo}_\mathrm{ex}(t)\rvert_{A_1, B_1 \ \mathrm{initial}} & \mathrm{for} \ t \in \big(0, \frac{T}{2}\big) \\
        \chi^\mathrm{lo}_\mathrm{co}(t)\rvert_{A_2, B_2 \ \mathrm{glued}} & \mathrm{for} \ t \in \big(\frac{T}{2}, T\big) \\
        \chi^\mathrm{lo}_\mathrm{ex}(t)\rvert_{A_1, B_1 \ \mathrm{glued}} & \mathrm{for} \ t \in \big(T, \frac{3T}{2}\big) \\
        \qquad\qquad \vdots & \quad\qquad\vdots 
    \end{cases} \label{eq:glued_mode_function}
\end{align}
The Floquet solutions are those that have initial $A_1, B_1$ such that
\begin{align}
    \chi_\mathrm{ex}^\mathrm{lo}(T) = e^{F T} \, \chi_\mathrm{ex}^\mathrm{lo}(0).
\end{align}
As long as $|e^{F T}| = 1$, the solution oscillates and does not grow or decay. It can be shown that two such solutions exist and that $|e^{F T}| = 1$ as long as
\begin{align}
    \frac{a_\mathrm{max}}{a_\mathrm{min}} < e^2 \approx 7.4,
\end{align}
which is always fulfilled experimentally due to the bound on the maximal scale factor of around 3. Therefore, in the long wavelength regime, there is no resonant growth of the mode functions.

\subsubsection*{Short Wavelength}

In the short wavelength regime, the solutions for the mode functions are
\begin{align}
    \chi_\mathrm{ex}^\mathrm{hi}(\tau) = \sqrt{a_\mathrm{min} + \alpha \tau} \left[ A_1 \cos\Big( \nu \log(a_\mathrm{min} + \alpha\tau ) \Big) + B_1 \sin\Big( \nu \log(a_\mathrm{min} + \alpha\tau) \Big) \right]
\end{align}
in the expanding phase and
\begin{align}
    \chi_\mathrm{co}^\mathrm{hi}(\tau) = \sqrt{a_\mathrm{min} + \alpha (T-\tau)} \left[ A_2 \cos\Big( \nu \log\big(a_\mathrm{min} + \alpha (T-\tau) \big) \Big) + B_2 \sin\Big( \nu \log\big(a_\mathrm{min} + \alpha (T-\tau)\big) \Big) \right]
\end{align}
in the contracting phase. This time, the parameter $\nu$ is given by
\begin{align}
    \nu = \sqrt{\frac{k^2}{\alpha^2} - \frac{1}{4}}. \label{eq:nudef}
\end{align}
Again, there are two Floquet solutions with
\begin{align}
    e^{\pm F T} &= \frac{1}{4\nu^2} \Bigg[ -1 + (1 + 4\nu^2) \cos\Big( 2\nu \log\frac{a_\mathrm{max}}{a_\mathrm{min}}\Big) \notag\\
    &\quad + \sqrt{-16\nu^2 + \Big( -1 + (1 + 4\nu^2) \cos\Big( 2\nu \log\frac{a_\mathrm{max}}{a_\mathrm{min}} \Big) \Big)^2} \ \Bigg].
\end{align}
The absolute value of these factors is more complicated than for long wavelengths and can become larger (or smaller) than $1$ for realistic values of $a_\mathrm{max}$ and $a_\mathrm{min}$. The condition for this growth is
\begin{align}
    \log\frac{a_\mathrm{max}}{a_\mathrm{min}} \in \Big[ \frac{\arctan(2\nu)}{\nu}, \frac{\pi - \arctan(2\nu)}{\nu} \Big] + \frac{\pi}{\nu} \mathbb{N}. \label{eq:floquet_growth_condition}
\end{align}
As expected for resonant fluctuations, there are infinitely many harmonics of a resonant window. Figure \ref{fig:4pictures} visualizes the resonant conditions and the corresponding Floquet factor $e^{F T}$ as a function of $\nu$. The figure shows clear resonant bands that would facilitate exponential growth of the mode function. Since the propagator is just the product of two mode functions,
\begin{align}
    \langle \phi_{\vec k}(t) \phi_{\vec k}(t) \rangle = G_k(t, t) = \phi_k(t) \phi_k^*(t),
\end{align}
this growth leads to resonant particle production. In a realistic system, our EFT breaks down when the fluctuations become too large and a steady state of particle production and dissipative effects takes over. This is precisely the phenomenon observed experimentally, for example in \cite{Liebster:2023mmf}. The referenced experiment uses a sine-wave scattering length $a_s(t)$, which is related to the scale factor via
\begin{align}
    a(t) \propto \frac{1}{\sqrt{a_s(t)}}
\end{align}
instead of a triangle wave. Therefore, the results are hard to compare quantitatively. Qualitatively, figure 1 of \cite{Liebster:2023mmf} shows one resonant band. Higher resonances most likely occur at very high $k$, such that they are not in the displayed range, or the EFT breaks down before the second resonant range.

\subsection{Iterative Integration}

As explained around \eqref{eq:glued_mode_function}, we can only write the mode functions in a piece-wise manner, with gluing conditions between the pieces. However, this is not a problem. We can solve the Schwinger-Keldysh time integral for every piece as a function of the coefficients $A_{1/2}$ and $B_{1/2}$, and then sum over pieces using the coefficients the gluing conditions give us. Since there seems to be no clear pattern for iterated gluing, the last step has to be performed numerically. This does obscure the exact phenomena somewhat, but in this case, it is possible to read them off from the plots, see figure \ref{fig:3_pt_periodic}.

We will focus on low-$k$ modes. The high $k$ modes can be treated similarly using the analytic continuation in $\mu \rightarrow i \mu = \nu$. Also, we will only present the derivation for the expanding phase of the scale factor, since the contracting phase is similar. The interaction term the BEC gives us in \eqref{eq:scale_free_action_final} involves time derivatives of the field $\phi_k(t)$. To compute the interactions, we need the integral
\begin{align}
    I_\mathrm{ex}^\mathrm{lo} &\coloneq \int\limits_0^{T/2} \d \tau \, a^2(\tau) \Bigg(\dot\phi_{k_1} \phi_{k_2} \phi_{k_3} \cdot \frac{k_1^2 - k_2^2 - k_3^2}{2} + \mathrm{permutations} \Bigg).
\end{align}
The derivatives act either on $\chi_k(t)$ or on $a(t)$. Acting on $a(t)$ gives a factor of $\alpha$. Acting on $\chi_k(t)$ also gives a factor of $\alpha$, and a factor of $\frac{1}{2} \pm \mu$, depending on if it hit the $A$ or the $B$ term in \eqref{eq:low_k_expanding_mode}. Lastly, it produces a factor of $(a_\mathrm{min} + \alpha \tau)^{-1}$, which is just $1/a(t)$ in the expanding phase. Counting powers of $a(t)$ for every generated term, the only relevant integral is
\begin{align}
    I(\Sigma \mu) \coloneq \int\limits_0^{T/2} \d \tau \, \big( a_\mathrm{min} + \alpha \tau\big)^{-\frac{1}{2} + \Sigma \mu},
\end{align}
where $\Sigma\mu = \pm \mu_1 \pm \mu_2 \pm \mu_3$, and the $\pm$ is decided by taking the $A$ or $B$ term. After some algebra, one finds
\begin{align}
    I_\mathrm{ex}^\mathrm{lo} &= \sum\limits_{\pm,\pm,\pm} \alpha \binom{A1}{B1}_{\mu_1} \binom{A2}{B2}_{\mu_2} \binom{A3}{B3}_{\mu_3} I(\pm \mu_1 \pm \mu_2 \pm \mu_3) \times \Bigg[\Big( \pm \mu_1 - \frac{1}{2} \Big) \frac{k_1^2-k_2^2-k_3^2}{2} \notag\\
    & \qquad +  \Big( \pm \mu_2 - \frac{1}{2} \Big) \frac{k_2^2-k_1^2-k_3^2}{2} +  \Big( \pm \mu_3 - \frac{1}{2} \Big) \frac{k_3^2-k_1^2-k_2^2}{2} \Bigg].
\end{align}
The parentheses $\binom{A}{B}_{\mu}$ are shorthand for "if $\mu > 0$ take $A$, else $B$".
The results of the iterated gluing and integration for some relevant parameters are shown in figure \ref{fig:3_pt_periodic}.

\bibliographystyle{JHEP}
\bibliography{references}

\end{document}